\documentclass[secnumarabic,amssymb, nofootinbib, aps, prd]{revtex4-2}
\usepackage{amsmath,amssymb}
\usepackage{graphicx}
\usepackage{multirow}
\usepackage{caption}
\usepackage{subcaption}
\usepackage{hyperref}
\usepackage{url}

\begin{document}

\title{Realistic neutron stars in the constant-Q branch of f(Q) gravity}%

\author{Debadri Bhattacharjee}
\email{debadriwork@gmail.com}
\affiliation{IUCAA Centre for Astronomy Research and Development (ICARD), Department of Physics, Cooch Behar Panchanan Barma University, Vivekananda Street, District: Cooch Behar, Pin: 736101, West Bengal, India}

\author{Pradip Kumar Chattopadhyay}
\email{pkc\_76@rediffmail.com}
\affiliation{IUCAA Centre for Astronomy Research and Development (ICARD), Department of Physics, Cooch Behar Panchanan Barma University, Vivekananda Street, District: Cooch Behar, Pin: 736101, West Bengal, India}

\author{P. H. R. S. Moraes}
\email{moraes.phrs@gmail.com}
\affiliation{N\'ucleo de Astrof\'isica, Universidade Cidade de S\~ao Paulo, Rua Galv\~ao Bueno 868, S\~ao Paulo 01506-000, Brazil}
%\date{December 2018}%
\begin{abstract}
	We investigate the hydrostatic equilibrium and global structure of neutron stars within symmetric teleparallel gravity, where gravitational interactions are mediated by the non-metricity scalar $Q$. Working within the metric-affine formulation, we adopt the regular, constant-$Q$ branch ($Q = Q_*$) of $f(Q)$ gravity, which preserves regularity at the stellar center while maintaining explicit dependence on the non-linear functional form through $f(Q_*)$ and its derivatives. We propose three novel non-linear $f(Q)$ extensions: (i) a Born-Infeld-inspired model, (ii) a Hu--Sawicki analogue, and (iii) a Tsujikawa-inspired model. Incorporating state-of-the-art realistic unified equations of state, namely GPPVA(DD2) and SLy4, we numerically solve the modified Tolman-Oppenheimer-Volkoff equations and evaluate stellar stability via the Harrison-Zel'dovich-Novikov criterion. The Born-Infeld and Hu-Sawicki analogues smoothly recover the general relativistic limit as their control parameters increase, predicting stellar radii for low-mass objects ($M < 2.0\,M_\odot$) that match observational constraints from X-ray binaries and GW170817; however, their maximum masses remain capped below $2.43\,M_\odot$ (DD2) and $2.06\,M_\odot$ (SLy4). Conversely, the Tsujikawa-inspired model significantly stiffens the effective gravitational coupling, accommodating massive compact objects up to $\sim 3.36\,M_\odot$ and offering a viable physical origin for the $2.59_{-0.09}^{+0.08}\,M_\odot$ secondary object in GW190814, albeit at the expense of overpredicting radii for typical low-mass neutron stars. These contrasting behaviours demonstrate a key phenomenological complementarity between maximum mass and radius bounds within the constant-$Q$ framework, highlighting the utility of multimessenger observations in constraining non-linear non-metricity gravity.
\end{abstract}
\maketitle
\tableofcontents

\section{Introduction}\label{sec1} 

Neutron stars are among the most studied astrophysical objects, especially since the first detection of gravitational waves emitted by them \cite{abbott/2017} and the subsequent birth of multimessenger astrophysics. The event reported in \cite{abbott/2017}, namely GW170817, is the first observation of gravitational waves emitted by a pair of inspiralling neutron stars\footnote{All the previous gravitational wave detections had binary black holes in coalescence as their progenitors \cite{abbott/2016,abbott/2016b,abbott/2017b,abbott/2018,abbott/2017c}.}. The fusion of such neutron stars is the main candidate for the progenitor system of short gamma ray bursts \cite{kouveliotou/1993}, therefore, such an electromagnetic counterpart was expected to be seen. In \cite{abbott/2017d}, it was reported the Fermi gamma ray burst monitor detection of a short gamma ray burst (named GRB 170817A), with a time delay of $\sim1.7$s with respect to the merger time of GW170817.

A persistent and fundamental question regarding neutron star physics is its composition. Unlike the case of white dwarfs, in which the pressure sustaining the star comes mainly from a gas of degenerated electrons \cite{chandrasekhar/1931}, since they interact electromagnetically, and quantum electrodynamics is extremely well tested, such a pressure can be calculated with high precision. In the interior of neutron stars, however, density can be many times larger than nuclear density. In this regime, matter is governed by strong nuclear force and we are still not sure of its composition. It can contain neutrons and protons, hyperons, meson condensate, deconfined quark matter etc. Terrestrial experiments do not directly reproduce this range of densities, so there are several plausible equations of state (see, for instance, \cite{baym/2019,somasundaram/2023,raithel/2019,lindblom/2018}).

Recently, several constraints have been put to the neutron star equation of state, especially from the astrophysics of gravitational waves and multimessenger astrophysics \cite{dietrich/2020,radice/2018,tan/2020,raithel/2019b,zhu/2018}.

Another field of physics that is currently also under intense investigation is {\it modified gravity}. Einstein's theory of General Relativity provides the most successful description of gravitational phenomena across a wide range of systems and scales \cite{will/2014}. Nevertheless, several open theoretical and observational questions motivate the exploration of possible extensions and modifications of the theory. Among this open questions, one could quote dark matter \cite{bertone/2005,bertone/2018} and the behaviour of gravity at cosmological \cite{raveri/2020,li/2012} and quantum scales \cite{rovelli/1998,hossenfelder/2013}.

It is intuitive to think that any modified theory of gravity should also be tested in the astrophysical stellar regime. On this regard, modified gravity, as a matter of fact, could also act as a tool to increase the maximum masses theoretically predicted for neutron stars \cite{astashenok/2014,quartuccio/2025}, as the law of gravity itself can change in such regimes of high compactification. Such an increasing in the maximum mass predicted for neutron stars is very welcome in view of observed massive pulsars \cite{antoniadis/2013,riley/2021,nice/2005} and, especially, the secondary object in GW190814 \cite{Abbott}, that could be a neutron star with $\sim2.6$M$_\odot$. 

%There is a plethora of modified gravity models currently proposed in the literature. 

A quite applicable and popular modified gravity model is the $f(R)$ gravity \cite{zhao/2014,feng/2026,kusakabe/2015}, with $f(R)$ representing a generic function of the Ricci scalar $R$ to substitute $R$ itself in the Einstein-Hilbert action. Although observations place stringent constraints on deviations from General Relativity, a range of modified gravity models remains viable. Beyond $f(R)$ gravity, several alternative frameworks have been explored in the literature (see References \cite{dvali/2001,lue/2004,obukhov/2003,konno/2007,barcelo/2003}, for instance).

In the present article, we are going to work with the so-called $f(Q)$ gravity \cite{Jimenez3}, with $Q$ being the non-metricity scalar. This theory is formulated within the framework of symmetric teleparallel geometry, in which the affine connection has vanishing curvature and torsion, while gravitational effects are encoded in non-metricity, which describes the failure of the connection to preserve the metric under parallel transport. The symmetric teleparallel equivalent of General Relativity is recovered for an appropriately normalized gravitational Lagrangian linear in $Q$, whereas replacing this linear dependence with a general function $f(Q)$ yields modified gravitational dynamics.

Cosmological models derived from the $f(Q)$ gravity formalism can be seen in \cite{frusciante/2021,atayde/2021} (also in \cite{Jimenez3}). Cosmography and energy conditions were derived in \cite{mandal/2020,mandal/2020b}, respectively. Static and spherically symmetric solutions in $f(Q)$ gravity were obtained in \cite{wang/2022}.

An important aspect of exploring modified gravitational theories is to investigate how different nonlinear extensions of the gravitational Lagrangian influence the equilibrium properties of compact objects. In this work, we consider three distinct functional forms of $f(Q)$, motivated by well-established approaches to constructing viable modifications of gravity, namely the Born-Infeld, Hu-Sawicki, and Tsujikawa prescriptions. The selection of these models is intended to examine different classes of nonlinear corrections to the non-metricity scalar and their implications for neutron star structure. First, inspired by the Born–Infeld approach \cite{Born,Born1}, we consider a square-root extension of the gravitational Lagrangian. The characteristic nonlinear structure of Born-Infeld-type theories provides a natural framework for investigating modifications involving a nonlinear, regularised dependence on the underlying geometric invariant. Its inclusion allows us to examine how such a nonlinear extension affects the effective gravitational coupling and the resulting stellar configurations, while retaining a smooth general relativistic limit. Second, motivated by the Hu-Sawicki \cite{Hu} construction originally developed in the context of cosmological $f(R)$ gravity, we introduce a rational-function correction to the linear $Q$ term. The characteristic feature of this prescription is its controlled nonlinear contribution, governed by a dimensionless amplitude. This structure provides an opportunity to investigate how the strength and functional suppression of nonlinear corrections influence the maximum mass and radius of neutron stars. Third, following the Tsujikawa prescription \cite{Tsujikawa}, we consider a hyperbolic-tangent correction to the gravitational Lagrangian. The bounded and asymptotically saturating nature of the hyperbolic-tangent function offers a distinct phenomenological realisation of nonlinear gravity, in which the magnitude of the correction is controlled by a single characteristic parameter. This model therefore allows us to explore the consequences of a saturating geometric correction for the effective gravitational interaction within compact stars. Importantly, all three prescriptions admit the linear form $f(Q)=Q$ in appropriate parameter limits, providing a common general relativistic reference against which their respective predictions can be assessed. Their comparative investigation consequently offers a useful means of examining how different nonlinear structures of the gravitational Lagrangian affect neutron star masses, radii, and stability within the constant-Q branch of $f(Q)$ gravity.

In the present article, we are going to investigate the structure of neutron stars within the constant-$Q$ branch of $f(Q)$ gravity, in which the non-metricity scalar is taken to be constant. We will formulate and numerically solve the corresponding hydrostatic equilibrium equations using realistic equations of state, examining how the chosen forms of $f(Q)$ and their parameters affect the mass–radius relation and the maximum stellar mass relative to General Relativity. We will also compare the resulting stellar configurations with observational mass and radius constraints to assess the compatibility of these models with neutron star observations. In this regard, Section~\ref{sec2} describes the fundamentals of $f(Q)$ gravity leading to the intricate forms of the connection parameters and the modified field equations. In this section, we also demonstrate the construction of TOV equations and the imposed boundary conditions within a static, spherically symmetric space-time. The solutions of TOV equations for the novel choices of $f(Q)$ models and realistic equations of state yield maximum mass-radius relations for neutron stars, which are analysed in Section~\ref{sec3}. The physical viability of the present framework has been addressed in Section~\ref{sec4}. Finally, we summarise the main findings in Section~\ref{sec5}.

\section{The $f(Q)$ formalism}\label{sec2}

General Relativity Theory admits different geometric descriptions. In its conventional formulation, gravity is associated with curvature, whereas its teleparallel and symmetric teleparallel equivalents describe gravity through torsion and non-metricity, respectively \cite{Jimenez1,Jimenez3}. Symmetric teleparallel geometry is characterized by a flat, torsion-free affine connection and generally nonzero non-metricity. This provides the geometric setting for $f(Q)$ gravity, whose gravitational lagrangian is a generic function of the non-metricity scalar $Q$ (see References \cite{Jimenez3,heisenberg/2024}).

In the present article, we use geometrized units $G=c=1$, define $\kappa=8\pi$ and adopt the metric signature $(-,+,+,+)$. Both Latin and Greek tensor indices run over the four spacetime coordinates. Parentheses denote symmetrisation with unit weight, for example $A_{(ij)}=(A_{ij}+A_{ji})/2$.

We consider the metric-affine structure $(\mathcal{M},\Gamma^{\alpha}{}_{ij},g_{ij})$, where $\mathcal{M}$ is a smooth four-dimensional manifold, $\Gamma^{\alpha}{}_{ij}$ is an independent affine connection and $g_{ij}$ is the metric tensor. The derivative $\nabla$ is associated with this affine connection, while a comma denotes a partial derivative. The curvature, torsion and non-metricity tensors are defined, respectively, as \cite{heisenberg/2024}

\begin{align}
	R^{\sigma}_{\alpha ij}&=\Gamma^{\sigma}_{j\alpha,i}-\Gamma^{\sigma}_{i\alpha,j}+\Gamma^{\sigma}_{i\beta}\Gamma^{\beta}_{j\alpha}-\Gamma^{\sigma}_{j\beta}\Gamma^{\beta}_{i\alpha}, \label{eq1} \\ \mathcal{T}^{\alpha}_{ij}&=\Gamma^{\alpha}_{ij}-\Gamma^{\alpha}_{ji},  \label{eq2}\\
	Q_{\alpha ij}&=\nabla_{\alpha}\,g_{ij}=\partial_{\alpha}g_{ij}-2\Gamma^{\eta}_{\alpha(i}g_{j)\eta}. \label{eq3}
\end{align}

Symmetric teleparallel geometry imposes $R^{\sigma}{}_{\alpha ij}=0$ and $\mathcal{T}^{\alpha}{}_{ij}=0$. These conditions refer to the independent affine connection and do not require the curvature of the Levi-Civita connection to vanish. Since $Q_{\alpha ij}$ is symmetric in its last two indices, it has two independent traces,
\[
Q_{\alpha}=g^{ij}Q_{\alpha ij},
\qquad
\tilde Q_{\alpha}=g^{ij}Q_{i\alpha j}.
\]
Together with the full non-metricity tensor, these traces yield five independent parity-even quadratic contractions. Their general linear combination is \cite{Heisenberg}

\begin{equation}
	Q=c_{1}Q_{\alpha\beta\gamma}Q^{\alpha\beta\gamma}+c_{2}Q_{\alpha\beta\gamma}Q^{\beta\alpha\gamma}+c_{3}Q_{\alpha}Q^{\alpha}+c_{4}\tilde{Q}_{\alpha}\tilde{Q}^{\alpha}+c_{5}Q_{\alpha}\tilde{Q}^{\alpha}, \label{eq4}
\end{equation}
where $c_i$, $i=1,\ldots,5$, are constant coefficients. For the $f(Q)$ formalism used below, they are fixed to the symmetric teleparallel equivalent of General Relativity (also known as STEGR) values,
\[
c_1=-\frac14,\qquad c_2=\frac12,\qquad c_3=\frac14,
\qquad c_4=0,\qquad c_5=-\frac12.
\]
This choice specifies the scalar $Q$ and its sign convention throughout the remainder of the paper; the subsequent equations do not assume arbitrary $c_i$.

The non-metricity conjugate, or superpotential, is the tensor

\begin{equation}
	P^{\alpha}_{ij}=\frac{1}{2}\frac{\partial\,Q}{\partial\,Q_{\alpha}{}^{ij}}, \label{eq5}
\end{equation}
where the partial derivative is taken at fixed metric. For the coefficients specified above, its explicit form is
\[
P^{\alpha}{}_{ij}
=-\frac14Q^{\alpha}{}_{ij}
+\frac12Q_{(i}{}^{\alpha}{}_{j)}
+\frac14\left(Q^{\alpha}-\tilde Q^{\alpha}\right)g_{ij}
-\frac14\delta^{\alpha}{}_{(i}Q_{j)},
\]
with $\delta^{\alpha}{}_{i}$ the Kronecker delta. It satisfies $Q=P_{\alpha ij}Q^{\alpha ij}$. 

The gravitational action linear in this scalar, with flatness and vanishing torsion imposed by Lagrange multipliers, is

\begin{equation}
	S=\int_{\mathcal{M}}\,d^{4}x\,\left(\frac{1}{16\pi}\sqrt{-g}\,Q+\eta^{\beta ij}_{\alpha}R^{\alpha}_{\beta ij}+\sigma_{\alpha}{}^{ij}\mathcal{T}^{\alpha}_{ij}\right)+S_{\mathrm M}, \label{eq6} 
\end{equation}
where $g=\det(g_{ij})$, and $\eta^{\beta ij}_{\alpha}$ and $\sigma_{\alpha}{}^{ij}$ are Lagrange multiplier tensor densities of weight $+1$ and antisymmetric in $i,j$. Varying them imposes the curvature and torsion constraints, respectively \cite{Jimenez1,Jimenez2}. The matter action is $S_{\mathrm M}=\int_{\mathcal M}d^4x\sqrt{-g}\,\mathcal L_{\mathrm M}$, where $\mathcal L_{\mathrm M}$ is the matter lagrangian. We assume minimal coupling to the metric and no direct dependence of the matter action on $Q$ or on the independent affine connection.

Replacing $Q$ in the gravitational Lagrangian with a function $f(Q)$ gives the action of $f(Q)$ gravity \cite{heisenberg/2024},

\begin{equation}
	S_{f(Q)}=\int_{\mathcal{M}}d^{4}x\,\left[\frac{1}{16\pi}\sqrt{-g}\,f(Q)+\eta^{\beta ij}_{\alpha}R^{\alpha}_{\beta ij}+\sigma_{\alpha}{}^{ij}\mathcal{T}^{\alpha}_{ij}\right]+S_{\mathrm M}. \label{eq7}
\end{equation}

The energy-momentum tensor is defined by variation of the matter action,

\begin{equation}
	T_{ij}=-\frac{2}{\sqrt{-g}}\frac{\delta\,S_{\mathrm M}}{\delta\,g^{ij}}. \label{eq8}
\end{equation}

Variation of Eq.(\ref{eq7}) with respect to $g^{ij}$ yields

\begin{equation}
	\frac{2}{\sqrt{-g}}\nabla_{\alpha}\left[\sqrt{-g}\,P^{\alpha}_{ij}\,f'(Q)\right]+f'(Q)\,q_{ij}-\frac{1}{2}f(Q)\,g_{ij}=8\pi T_{ij}. \label{eq9}
\end{equation}
Our notation is such that $f'(Q)\equiv df/dQ$, $f''(Q)\equiv d^2f/dQ^2$ and $f^{'''}(Q)\equiv d^3f/dQ^3$. The symmetric tensor $q_{ij}$ is defined explicitly by \cite{Heisenberg} % The alternative notation $f_Q=f'(Q)$ and $f_{QQ}=f''(Q)$ will also be used when discussing the models.

\[
q_{ij}=P_{(i|\alpha\beta}Q_{j)}{}^{\alpha\beta}
-2P^{\alpha\beta}{}_{(j}Q_{\alpha\beta|i)},
\]
where vertical bars exclude enclosed indices from symmetrisation. For the STEGR coefficients specified above, Eq.(\ref{eq9}) can equivalently be written as

\begin{equation}
	\tilde{G}_{ij}=f'(Q)\,G_{ij}-\frac{1}{2}g_{ij}\left[f(Q)-f'(Q)Q\right]+2f''(Q)P^{\alpha}_{ij}(\partial_{\alpha}Q)=8\pi T_{ij}, \label{eq10}
\end{equation}
Here, $G_{ij}=\mathring R_{ij}-\tfrac12g_{ij}\mathring R$ is the Einstein tensor constructed from the Levi-Civita connection, and $\tilde G_{ij}$ denotes the complete geometric left-hand side of the modified metric equation. The symbols $\mathring R_{ij}$ and $\mathring R$ are the corresponding Ricci tensor and scalar, with $\mathring R_{ij}=\mathring R^{\alpha}{}_{i\alpha j}$. For $f(Q)=Q$, Eq.(\ref{eq10}) reduces to $G_{ij}=8\pi T_{ij}$.

Under the assumed matter coupling, $\delta S_{\mathrm M}/\delta\Gamma^{\alpha}{}_{ij}=0$, so that the matter hypermomentum, which measures the response to the independent connection, vanishes. Variation with respect to the connection and elimination of the Lagrange multipliers then yield

\begin{equation}
	\nabla_{i}\nabla_{j}
	\left(\sqrt{-g}\,f'(Q)P^{ij}{}_{\alpha}\right)=0.
	\label{eq11}
\end{equation}
In this expression, $P^{ij}{}_{\alpha}=g^{j\beta}P^i{}_{\beta\alpha}$ and the covariant derivatives act on a tensor density. Diffeomorphism invariance relates the metric and connection equations. On solutions of the field equations, the conservation statement needed below is

\begin{equation}
	\mathring{\nabla}_{i}\tilde{G}^{i}{}_{j}
	=8\pi\mathring{\nabla}_{i}T^{i}{}_{j}=0.
	\label{eq12}
\end{equation}
The derivative $\mathring\nabla$ is associated with the Levi-Civita connection. Thus the minimally coupled matter sector obeys the usual covariant conservation law.

In a metric-affine geometry, the connection can be decomposed as

\begin{equation}
	\Gamma^{\alpha}_{ij}=\{^{\alpha}_{ij}\}+\mathcal{K}^{\alpha}_{ij}+L^{\alpha}_{ij}, \label{eq13}
\end{equation}
where $\{^{\alpha}_{ij}\}$ is the Levi-Civita connection, $\mathcal K^{\alpha}{}_{ij}$ is the contorsion and $L^{\alpha}{}_{ij}$ is the disformation. Explicitly,
\begin{align*}
	\{^{\alpha}_{ij}\}
	&=\frac12g^{\alpha\beta}
	\left(\partial_i g_{\beta j}+\partial_jg_{\beta i}-\partial_\beta g_{ij}\right),\\
	\mathcal K^{\alpha}{}_{ij}
	&=\frac12\mathcal T^{\alpha}{}_{ij}+\mathcal T_{(i}{}^{\alpha}{}_{j)},
	\qquad
	L^{\alpha}{}_{ij}
	=\frac12Q^{\alpha}{}_{ij}-Q_{(i}{}^{\alpha}{}_{j)}.
\end{align*}
The contorsion vanishes in the torsion-free geometry considered here. Locally, a flat, torsion-free connection admits the representation

\begin{equation}
	\Gamma^{\alpha}_{ij}=\frac{\partial\,x^{\alpha}}{\partial\,\zeta^{\eta}}\partial_{i}\partial_{j}\zeta^{\eta}, \label{eq14}
\end{equation}
where $\zeta^{\eta}(x)$ are four functions with an invertible Jacobian and $\partial x^{\alpha}/\partial\zeta^{\eta}$ is the inverse Jacobian \cite{Jimenez3}. Equation~(\ref{eq14}) relies on the flatness and torsion-free conditions and does not represent an unrestricted affine connection.

For static, spherically symmetric configurations, we use coordinates $(t,r,\theta,\phi)$, where $r$ is the areal radius, with $g_{tt}=g_{tt}(r)$, $g_{rr}=g_{rr}(r)$, $g_{\theta\theta}=r^2$ and $g_{\phi\phi}=r^2\sin^2\theta$. We follow the spherical connection branch employed in Ref.~\cite{Heisenberg}. The radial connection components entering the equations below are $\Gamma^r_{rr}(r)$ and $\Gamma^r_{\theta\theta}(r)$, with the flatness relation
\[
\partial_r\Gamma^r_{\theta\theta}
=-1-\Gamma^r_{rr}\Gamma^r_{\theta\theta}.
\]
For this branch, the expression for $Q$ can be solved algebraically for $\Gamma^r_{rr}$, allowing us to use $(g_{tt},g_{rr},Q)$ together with the remaining connection relation. The non-metricity scalar is

\begin{eqnarray}
	Q=&\frac{1}{r^{2}g_{rr}^{2}g_{tt}(\Gamma^{r}_{\theta\theta})^{2}}\Bigg\{g_{tt}\Big[\Gamma^{r}_{\theta\theta}((\Gamma_{\theta\theta}^{r})^{2}g_{rr}-r^{2})(\partial_{r}g_{rr})+2g_{rr}\Big((\Gamma_{\theta\theta}^{r})^{2}-g_{rr}\Gamma^{r}_{rr}(\Gamma^{r}_{\theta\theta})^{3}+r^{2}+r\Gamma^{r}_{\theta\theta}\nonumber\\&(r\Gamma^{r}_{rr}+2)\Big)\Big]+\Gamma^{r}_{\theta\theta}g_{rr}[(\Gamma^{r}_{\theta\theta})^{2}g_{rr}+2r\Gamma^{r}_{\theta\theta}+r^{2}](\partial_{r}g_{tt})\Bigg\}. \label{eq15}
\end{eqnarray}

The $rr$ and $tt$ components of Eq.(\ref{eq10}) yield equations for $\partial_rg_{tt}$ and $\partial_rg_{rr}$, respectively,

\begin{equation}
	\partial_{r} g_{tt}
	=
	\frac{r g_{tt}8\pi}{f'(Q)}T_{rr}
	-
	\frac{g_{tt}\left[
		2+g_{rr}\left(Qr^{2}-2-\frac{f(Q)r^{2}}{f'(Q)}\right)
		\right]}{2r}
	+
	\frac{g_{tt}\left[
		r^{2}-g_{rr}\left(\Gamma^{r}_{\theta\theta}\right)^{2}
		\right]}
	{\Gamma^{r}_{\theta\theta}r f'(Q)}
	(\partial_{r} Q)f''(Q) \label{eq16}
\end{equation}
and

\begin{equation}
	\partial_{r} g_{rr}
	=
	-\frac{r g_{rr}^{2}8\pi}{f'(Q)g_{tt}}T_{tt}
	+
	\frac{g_{rr}\left[
		2+g_{rr}\left(Qr^{2}-2-\frac{f(Q)r^{2}}{f'(Q)}\right)
		\right]}{2r}
	+
	\frac{g_{rr}\left[(\Gamma^{r}_{\theta\theta})^{2}g_{rr}+2r\Gamma^{r}_{\theta\theta}+r^{2}\right]}{\Gamma^{r}_{\theta\theta}r f'(Q)}
	(\partial_{r} Q)f''(Q). \label{eq17}
\end{equation}
Here, $T_{tt}$ and $T_{rr}$ are covariant coordinate components of the energy-momentum tensor. In vacuum, these equations reduce to the corresponding expressions obtained in the black-hole analysis of Ref.~\cite{DAmbrosio3}, providing a consistency check of the reduction. Solving Eq.(\ref{eq15}) for $\Gamma^r_{rr}$ and substituting Eqs.(\ref{eq16}) and (\ref{eq17}) gives

\begin{align}
	\Gamma^{r}_{rr}&=\frac{\Gamma^{r}_{\theta\theta} f(Q) g_{rr} r^{2}\left(\Gamma^{r}_{\theta\theta}+r\right)+\left\{2r^{2}
		+\Gamma^{r}_{\theta\theta}
		\left[
		2r
		+
		g_{rr}
		\left(
		2\left(r+\Gamma^{r}_{\theta\theta}\right)
		-r^{2}Q
		\left(r+2\Gamma^{r}_{\theta\theta}\right)
		\right)
		\right]
		\right\}
		f'(Q)
	}{
		2f'(Q)\Gamma^{r}_{\theta\theta}
		\left[
		\left(\Gamma^{r}_{\theta\theta}\right)^{2}g_{rr}
		-r^{2}
		\right]
	}
	\nonumber\\&-
	\frac{r8\pi}{2f'(Q)}
	\left\{
	\frac{g_{rr}}{g_{tt}}T_{tt}
	-
	\frac{
		\left(\Gamma^{r}_{\theta\theta}\right)^{2}g_{rr}
		+2r\Gamma^{r}_{\theta\theta}
		+r^{2}
	}{
		\left(\Gamma^{r}_{\theta\theta}\right)^{2}g_{rr}
		-r^{2}
	}
	T_{rr}
	\right\},\label{eq18}
\end{align}
so that the flatness relation becomes

\begin{align}
	\partial_r\Gamma^{r}_{\theta\theta}&
	=
	-\frac{
		\Gamma^{r}_{\theta\theta}
		\left\{
		f(Q)g_{rr}r^{2}
		\left(\Gamma^{r}_{\theta\theta}+r\right)
		+
		\left[
		2r
		-g_{rr}
		\left(2\Gamma^{r}_{\theta\theta}+r\right)
		\left(Qr^{2}-2\right)
		\right]
		f'(Q)
		\right\}
	}{
		2f'(Q)
		\left[
		\left(\Gamma^{r}_{\theta\theta}\right)^{2}g_{rr}
		-r^{2}
		\right]
	}
	\nonumber\\&+
	\frac{
		r8\pi\Gamma^{r}_{\theta\theta}
	}{
		2f'(Q)
	}
	\left\{
	\frac{g_{rr}}{g_{tt}}T_{tt}
	-
	\frac{
		\left(\Gamma^{r}_{\theta\theta}\right)^{2}g_{rr}
		+2r\Gamma^{r}_{\theta\theta}
		+r^{2}
	}{
		\left(\Gamma^{r}_{\theta\theta}\right)^{2}g_{rr}
		-r^{2}
	}
	T_{rr}
	\right\}. \label{eq19}
\end{align}

Equation~(\ref{eq18}) reconstructs $\Gamma^r_{rr}$ in terms of the chosen variables, while Equation (\ref{eq19}) determines the remaining radial connection function. Using the connection equation, or the equivalent conservation relation between the field equations, gives

\begin{align}
	\partial_r^2 Q
	=&
	-\frac{r\,8\pi(\partial_r Q)}
	{2f'(Q)}
	\Bigg\{
	T_{rr}
	-
	\frac{
		g_{rr}\Big[
		\big(\Gamma^r_{\theta\theta}\big)^2g_{rr}
		+r^2
		\Big]
	}{
		g_{tt}\Big[
		\big(\Gamma^r_{\theta\theta}\big)^2g_{rr}
		-r^2
		\Big]
	}
	T_{tt}
	\Bigg\}+
	\frac{\partial_r Q}
	{2\Big[
		\big(\Gamma^r_{\theta\theta}\big)^2g_{rr}
		-r^2
		\Big]}
	\Bigg\{
	2r
	+
	g_{rr}\Bigg[
	4\Gamma^r_{\theta\theta}\nonumber\\&
	\Big(-\partial_r\Gamma^r_{\theta\theta}\Big)
	+2r
	-Qr^3
	\Bigg]
	+
	g_{rr}r^3\frac{f(Q)}{f'(Q)}
	-2\Bigg[
	\big(\Gamma^r_{\theta\theta}\big)^2g_{rr}
	+2\Gamma^r_{\theta\theta}g_{rr}r
	+r^2
	\Bigg]
	\frac{f''(Q)}{f'(Q)}(\partial_r Q)
	\nonumber\\&
	-2\Bigg[
	\big(\Gamma^r_{\theta\theta}\big)^2g_{rr}
	-r^2
	\Bigg]
	\frac{f^{(3)}(Q)}{f''(Q)}(\partial_r Q)
	\Bigg\}. \label{eq20}
\end{align}
These rearranged expressions apply where their denominators do not vanish. In particular, the equations for the metric functions require $f'(Q)\ne0$, while the displayed equation for $\partial_r^2Q$ also assumes $f''(Q)\ne0$. Degenerate cases, including the linear General Relativity limit, must be treated directly with the field equations before division by these factors.

\subsection{Spherically symmetric perfect fluid configuration and boundary constraints}

We write the static, spherically symmetric line element as

\begin{equation}
	ds^{2}=-e^{\nu(r)}\,dt^{2}+e^{\lambda(r)}\,dr^{2}+r^{2}(d\theta^{2}+\sin^{2}\theta\,d\phi^{2}),
	\label{eq21} 
\end{equation}
where $\nu(r)$ and $\lambda(r)$ are the metric potentials, so that $g_{tt}=-e^{\nu(r)}$ and $g_{rr}=e^{\lambda(r)}$. For an isotropic perfect fluid,

\begin{equation}
	T_{ij}=(\rho+p)u_{i}u_{j}+pg_{ij},  \label{eq22}
\end{equation}
where $\rho(r)$ is the energy density in the fluid rest frame, $p(r)$ is the pressure and $u_i=g_{ij}u^j$ is the covariant four-velocity, normalized by $u^iu_i=-1$. A static fluid has $u^i=(e^{-\nu/2},0,0,0)$, giving
\[
T_{tt}=\rho e^{\nu},\qquad T_{rr}=pe^{\lambda},\qquad
T_{\theta\theta}=pr^2,\qquad T_{\phi\phi}=pr^2\sin^2\theta.
\]

We introduce the geometric mass function $m(r)$ by $e^{-\lambda(r)}=1-2m(r)/r$. This definition is a parameterization of the radial metric; identifying a Schwarzschild exterior additionally requires the appropriate vacuum solution and matching conditions. The symbol $m(r)$ denotes the mass function, whereas the constant $m$ introduced in the Hu--Sawicki model in Section \ref{sec3} is a model exponent. %, or equivalently

%\begin{equation}
%    \lambda(r)=\ln\Bigg[1-\frac{2m(r)}{r}\Bigg]^{-1}. \label{eq23}
%\end{equation}

The radial component of $\mathring\nabla_iT^i{}_j=0$ gives hydrostatic equilibrium,

\begin{equation}
	\frac{d\nu}{dr}=-\frac{2}{\rho+p}\frac{dp}{dr}. \label{eq24}
\end{equation}
or, without division by $\rho+p$, $dp/dr=-(\rho+p)(d\nu/dr)/2$. The Levi-Civita connection is determined by the metric alone. Under the minimal-coupling assumption, Eq.~(\ref{eq24}) therefore remains valid independently of the choice of $f(Q)$ and of the compatible affine connection \cite{heisenberg/2024}.

Substituting the metric and perfect-fluid expressions into the radial field equations yields the stellar structure system \cite{Heisenberg},

\begin{equation}
	\frac{dm}{dr}=\frac{r^{2}}{2f'(Q)}
	\left\{-\frac{1}{2}\left[f(Q)-f'(Q)Q\right]+
	\frac{\left(\Gamma^{r}_{\theta\theta}+r\right)^{2}
		-2m\left(2\Gamma^{r}_{\theta\theta}+r\right)}
	{r^{2}\Gamma^{r}_{\theta\theta}}(\partial_{r}Q)f''(Q)+8\pi\rho\right\}, \label{eq25}
\end{equation}

\begin{equation}
	\frac{dp}{dr}=-\frac{p+\rho}{2f'(Q)(r-2m)}\left\{
	\frac{r^{3}f(Q)+f'(Q)\left(4m-r^{3}Q\right)}
	{2r}-\frac{\left(\Gamma^{r}_{\theta\theta}\right)^{2}
		+2rm-r^{2}}{\Gamma^{r}_{\theta\theta}}(\partial_{r}Q)f''(Q)+r^{2}8\pi p\right\}, \label{eq26}
\end{equation}

\begin{align}
	\partial_{r}^{2}Q=&
	\frac{\partial_{r}Q}
	{2\left[
		\left(\Gamma^{r}_{\theta\theta}\right)^{2}
		+2rm-r^{2}
		\right]}
	\Bigg\{
	4(r-m)
	+4\Gamma^{r}_{\theta\theta}
	\left(-\partial_{r}\Gamma^{r}_{\theta\theta}\right)
	-Qr^{3}
	+r^{3}\frac{f(Q)}{f'(Q)}\nonumber
	\\
	&
	-2\left[
	\left(\Gamma^{r}_{\theta\theta}+r\right)^{2}
	-2rm
	\right]
	\frac{f''(Q)}{f'(Q)}
	(\partial_{r}Q)
	-2\left[
	\left(\Gamma^{r}_{\theta\theta}\right)^{2}
	+2rm-r^{2}
	\right]
	\frac{f^{(3)}(Q)}{f''(Q)}
	(\partial_{r}Q)
	\Bigg\}
	\nonumber\\
	&-
	\frac{r^{2}\kappa(\partial_{r}Q)}
	{2f'(Q)(r-2m)}
	\left\{
	p+
	\frac{
		\left(\Gamma^{r}_{\theta\theta}\right)^{2}
		-2rm+r^{2}
	}{
		\left(\Gamma^{r}_{\theta\theta}\right)^{2}
		+2rm-r^{2}
	}\rho
	\right\}, \label{eq27}
\end{align}

\begin{align}
	\partial_{r}\Gamma^{r}_{\theta\theta}
	=&
	-\frac{
		\Gamma^{r}_{\theta\theta}
		\left\{
		f(Q)r^{2}
		\left(\Gamma^{r}_{\theta\theta}+r\right)
		+
		\left[
		2(r-2m)
		-\left(2\Gamma^{r}_{\theta\theta}+r\right)
		\left(Qr^{2}-2\right)
		\right]f'(Q)
		\right\}
	}{
		2f'(Q)
		\left[
		\left(\Gamma^{r}_{\theta\theta}\right)^{2}
		+2rm-r^{2}
		\right]
	}
	\nonumber\\
	&-
	\frac{
		r^{2}\kappa\Gamma^{r}_{\theta\theta}
	}{
		2f'(Q)(r-2m)
	}
	\left\{
	\rho+
	\frac{
		\left(\Gamma^{r}_{\theta\theta}+r\right)^{2}
		-2m\left(2\Gamma^{r}_{\theta\theta}+r\right)
	}{
		\left(\Gamma^{r}_{\theta\theta}\right)^{2}
		+2rm-r^{2}
	}
	p
	\right\}. \label{eq28}
\end{align}

For a general non-constant-$Q$ configuration, Eqs.(\ref{eq25})--(\ref{eq28}) must be supplemented by an equation of state (EoS), relating $p$ and $\rho$, and suitable boundary data. The same restrictions on the denominators apply, including those involving $\Gamma^r_{\theta\theta}$, $r-2m$ and $(\Gamma^r_{\theta\theta})^2+2rm-r^2$.

To discuss central regularity, we write the local expansions
\begin{align*}
	p(r)&=p^{(0)}+p^{(1)}r+p^{(2)}r^2+\mathcal O(r^3),\\
	\rho(r)&=\rho^{(0)}+\rho^{(1)}r+\rho^{(2)}r^2+\mathcal O(r^3),\\
	m(r)&=m^{(0)}+m^{(1)}r+m^{(2)}r^2+m^{(3)}r^3+\mathcal O(r^4),
\end{align*}
The superscripts in these expansions label coefficients. A smooth spherical centre has finite $p^{(0)}$ and $\rho^{(0)}$, vanishing first radial derivatives of smooth scalar profiles and $m(r)=\mathcal O(r^3)$. The EoS fixes the relation between the central pressure and density. The stellar surface is defined by $p(R)=0$, where $R$ is the stellar radius. The exterior and the connection must satisfy the corresponding vacuum and matching conditions.

The regularity conditions at the stellar centre require the metric functions and matter variables to remain finite, together with the vanishing of the first radial derivatives of smooth scalar quantities. In particular, the non-metricity scalar admits a regular expansion of the form
\begin{align*}
	Q(r)&=Q^{(0)}+Q^{(1)}r+Q^{(2)}r^2+\mathcal O(r^3).
\end{align*}
It is important to note that central regularity alone does not necessarily require the higher-order coefficients to vanish and, therefore, does not uniquely imply a radially constant non-metricity scalar. Motivated by the regularity analysis of Ref.\cite{Heisenberg}, the present study focuses on the constant-$Q$ branch. We denote its constant value by $Q_\star$, so that $Q(r)=Q_\star$ and $\partial_rQ=0$. This notation distinguishes the branch value from the parameter $Q_0$ appearing in the models of Sec.\ref{sec3}; the two are not identified by definition. The restriction to this branch is part of the setting considered here, rather than a general consequence of spherical symmetry alone. Its consistency must be understood together with the corresponding connection field equation and the interior-exterior matching conditions. Accordingly, the following stellar structure analysis is confined to configurations for which the prescribed constant value of $Q$ is compatible with these requirements.

On this branch, the explicit derivative terms in Eqs.(\ref{eq25}) and (\ref{eq26}) vanish. For $f'(Q_\star)\ne0$, the equations used for hydrostatic equilibrium become
\begin{align*}
	\frac{dm}{dr}
	&=\frac{r^2}{2f'(Q_\star)}
	\left\{8\pi\rho-\frac12\left[f(Q_\star)-Q_\star f'(Q_\star)\right]\right\},\\
	\frac{dp}{dr}
	&=-\frac{\rho+p}{r(r-2m)}
	\left\{m+\frac{4\pi r^3p}{f'(Q_\star)}
	+\frac{r^3\left[f(Q_\star)-Q_\star f'(Q_\star)\right]}{4f'(Q_\star)}\right\}.
\end{align*}
These expressions retain the dependence on the chosen model through $f(Q_\star)$ and $f'(Q_\star)$. In Eq.(\ref{eq10}), the same reduction gives an Einstein-type metric equation with a rescaled matter coupling and a constant term. The connection must still realise the prescribed value $Q_\star$ consistently. Thus, restricting to constant $Q$ does not, by itself, require all stellar predictions to coincide with those of General Relativity with the original coupling.

For $f(Q)=Q$, the equations reduce to the standard Tolman--Oppenheimer--Volkoff (TOV) system \cite{Tolman,Oppenheimer}.
%\[
%\frac{dm}{dr}=4\pi r^2\rho,\qquad
%\frac{dp}{dr}=-\frac{(\rho+p)(m+4\pi r^3p)}{r(r-2m)}.
%\]
%%%%%%%%%%%%%%%%%%%%%%%%%%%%%%%%%%%%%%%%%%%%%
\subsection{Effective gravitational interpretation of the constant-Q branch} 
The structure of Eqs. (\ref{eq24}) and (\ref{eq25}) can be further understood by introducing the following effective quantities:
\begin{equation}
	G_{\rm eff}=\frac{1}{f'(Q_\star)},
	\qquad
	\Lambda_{\rm eff}=-\frac{f(Q_\star)-Q_\star f'(Q_\star)}{2f'(Q_\star)}. \label{eq26a}
\end{equation}
In terms of these definitions, the modified TOV equations can be expressed as
\begin{equation}
	\frac{dm}{dr}
	=4\pi G_{\rm eff}\rho r^2
	+\frac{\Lambda_{\rm eff}}{2}r^2,~~~
	\text{and}
	~~~
	\frac{dp}{dr}
	=-\frac{\rho+p}{r(r-2m)}
	\left[
	m+4\pi G_{\rm eff}pr^3
	-\frac{\Lambda_{\rm eff}}{2}r^3
	\right]. \label{eq26b}
\end{equation}
These expressions reveal that, within the constant-$Q$ branch, the influence of the underlying gravitational Lagrangian on the stellar equilibrium equations is encoded entirely in the two quantities $G_{\rm eff}$ and $\Lambda_{\rm eff}$, evaluated at the constant value $Q_\star$. The first quantity characterises the effective coupling between matter and the gravitational sector, whereas the second accounts for the additional constant contribution arising from the combination $f(Q_\star)-Q_\star f'(Q_\star)$. Consequently, although different functional forms of $f(Q)$ may lead to different stellar configurations, their effects on the background equilibrium equations are determined by the corresponding values of these two effective quantities. In particular, two distinct functional forms can yield identical stellar equilibrium equations if they produce the same $G_{\rm eff}$ and $\Lambda_{\rm eff}$. This provides a useful framework for comparing the predictions of different nonlinear $f(Q)$ models and identifying the extent to which their effects on compact-star properties can be distinguished. Furthermore, in the limit $f(Q)\rightarrow Q$, one obtains $G_{\rm eff}\rightarrow 1$ and $\Lambda_{\rm eff}\rightarrow 0$, thereby recovering the standard general relativistic TOV equations. It should be emphasised that this effective description concerns the background stellar structure on the constant-Q branch and does not, by itself, establish the equivalence of the complete metric-affine theory to general relativity.

In the following section, we apply the constant-$Q$ reduction of Eqs.(\ref{eq25}) and (\ref{eq26}) to the three proposed forms of $f(Q)$, using the DD2 \cite{Grill,Typel,Pearson} and SLy4 EoSs \cite{Gulminelli,Chabanat,Danielewicz}.

\section{Novel $f(Q)$ models and realistic EoS}\label{sec3} 

The present paper is intended to showcase the impact of the constant-Q branch of $f(Q)$ gravity on the global properties of neutron stars. We consider the constant-$Q$ branch because it simplifies the
stellar structure equations while preserving their dependence
on the chosen gravitational model through $f(Q)$ and $f'(Q)$. For constant $Q$, the
explicit radial derivative terms vanish, simplifying the stellar
structure equations, whereas $f(Q)$ and $f'(Q)$ still enter the
effective gravitational coupling and the constant term in the
metric field equations. This allows us to compare the three
models using realistic EoSs before addressing the more general
case of radially varying non-metricity. The constant-$Q$
restriction is therefore adopted as a simplifying assumption;
regularity and compatibility with the connection equations
and boundary conditions must still be imposed.

In this regard, we introduce three novel forms of $f(Q)$ models that incorporate non-linear contributions from the non-metricity $Q$ while modifying the TOV equations. The newly introduced $f(Q)$ models are as follows: (i) Born-Infeld inspired model \cite{Born,Born1}, (ii) Hu-Sawicki analogue \cite{Hu}, and (iii) Tsujikawa-inspired model \cite{Tsujikawa}. Furthermore, to study neutron stars, we employ realistic state-of-the-art EoSs, {\it viz.} GPPVA(DD2) \cite{Grill,Typel,Pearson}\footnote{Following \url{https://compose.obspm.fr}, GPPVA(DD2) describes a neutron star unified inner crust-core EoS and yields a maximum mass of 2.42~$\mathrm{M_{\odot}}$. GPPVA(DD2) remains the stiffest EoS for a realistic neutron star, to date.} and SLy4 \cite{Gulminelli,Chabanat,Danielewicz}\footnote{Following \url{https://compose.obspm.fr}, SLy4 describes a unified EoS describing the interior cold and dense neutron star matter in beta equilibrium and yields a maximum mass of 2.05-2.06~$\mathrm{M_{\odot}}$}. 

Notably, throughout this work, we adopt geometrised units with $G=c=1$. Accordingly, the non-metricity scalar $Q$, the gravitational Lagrangian $f(Q)$ carry dimensions of inverse length squared. Using these proposed forms of $f(Q)$ models and the realistic EoSs, we solve the modified TOV equations for the constant-Q branch and study the maximum mass-radius relationship for each model. The obtained results are further compared with observational constraints to assess the feasibility of the proposed $f(Q)$ models.

\subsection{Born-Infeld inspired model: $f(Q)=Q_{0}\left[\sqrt{1+\frac{2Q}{Q_{0}}}-1\right]$}

The classes of theories that are generally described under the Born-Infeld umbrella share a common feature of incorporating a square root term intended to regularise the divergences. The pioneering work of Born and Infeld \cite{Born,Born1} originated from the aspect of finiteness, according to which a bounded physical parameter can never be infinite. In a more generalised way, they argued that the square root term must emerge from symmetry. In this regard, Born and Infeld proposed that the Lorentz symmetry of Maxwell's theory must be extended in their new approach. Considering a complete set of coordinate transformations as the initiator of the new symmetry, which recovered Maxwell's theory under appropriate limits, a new non-linear theory of electromagnetism was formulated and effectively termed the Born-Infeld electromagnetism \cite{Born,Born1}. Given the immense success of Born-Infeld theory, it was effectively applied to resolve the problems of GR. The first, yet futile, attempt was made by Deser and Gibbons \cite{Deser}. A direct implementation of Born-Infeld theory coupled with Ricci tensor produces ghost degrees of freedom pertaining to Ostrogradski instability associated with higher-order equations of motion produced by the theory \cite{Woodard,Woodard1}. The problem of the ghost field was resolved by the approach considered by Vollick \cite{Vollick}. In this work, the action was formulated within the metric-affine framework, in which the connection is not constrained by the metric but is instead regarded as an independent field. This framework can be effectively realised as a combination of the original affine theory of gravity proposed by Eddington and the Born-Infeld theory. Later, Ba\~nados and Ferreira \cite{Banados} modified the method outlined by Vollick, and it is referred to as Eddington-inspired Born-Infeld gravity (EiBI). In this framework, Ba\~nados and Ferreira \cite{Banados} explored the non-singular black holes and cosmological solutions. Based on the Born-Infeld framework, we introduce the model,
\begin{equation}
	f(Q)=Q_{0}\left[\sqrt{1+\frac{2Q}{Q_{0}}}-1\right]. \label{eq3.1}
\end{equation} 
Here, $Q_{0}$ is termed the control parameter. The numerical values of $Q_0$ employed in the present analysis are expressed in the same units as the non-metricity scalar, ensuring dimensional consistency in the adopted gravitational model. This model may be applied to the case of realistic neutron stars as it may soften the gravitational singularities. Moreover, recent astrophysical observations can be employed to constrain the parameter space of the EiBI theory of gravity. In this regard, several significant degeneracies arise which make it difficult to distinguish the contributions of EiBI theory from GR. The use of Born-Infeld theory in this context may help in understanding whether these degeneracies are entirely from EiBI theory or common to a large class of gravitational theories.  

The following analysis addresses the maximum mass-radius relationship for realistic neutron stars as obtained by using Eq.~(\ref{eq3.1}) in the modified TOV equations along with the realistic EoSs.   

\begin{figure}[t!]
	\begin{minipage}{0.45\textwidth}
		\hspace{-1cm}
		\includegraphics[width=8cm]{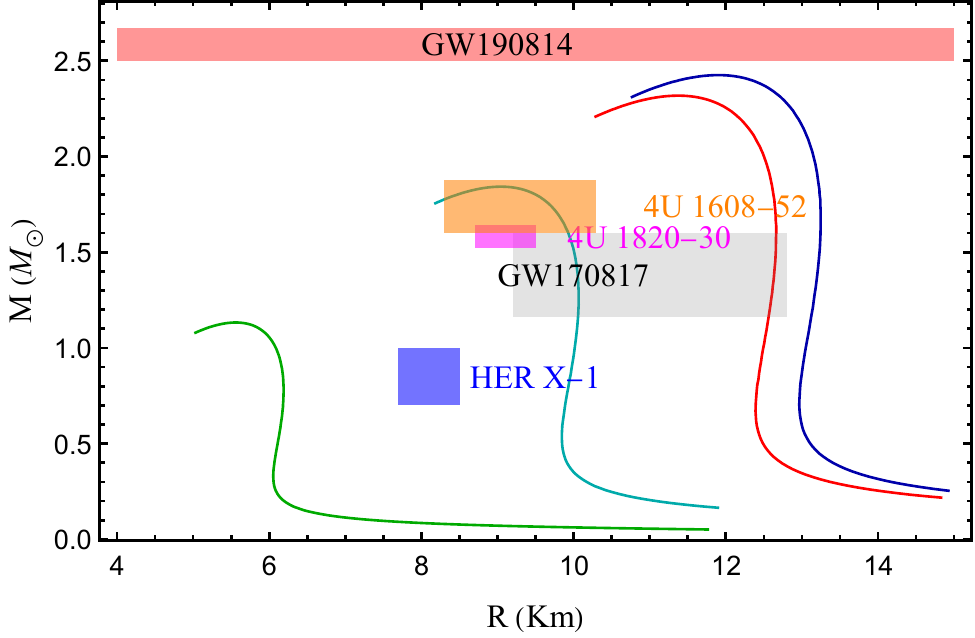}
		\subcaption{Maximum mass-radius plots using DD2 EoS.}
		\label{fig1} 
	\end{minipage}
	\hfil
	\begin{minipage}{0.45\textwidth}
		\hspace{-0.5cm}
		\includegraphics[width=8cm]{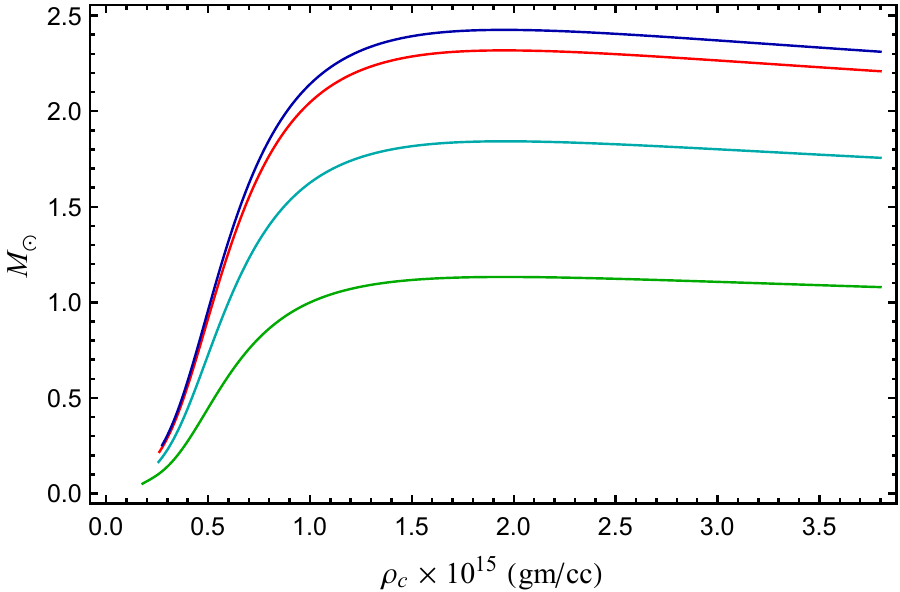}
		\subcaption{Mass vs Central density $(\rho_{c})$ using DD2 EoS.}
		\label{fig2} 
	\end{minipage}
	\begin{minipage}{0.45\textwidth}
		\hspace{-1cm}
		\includegraphics[width=8cm]{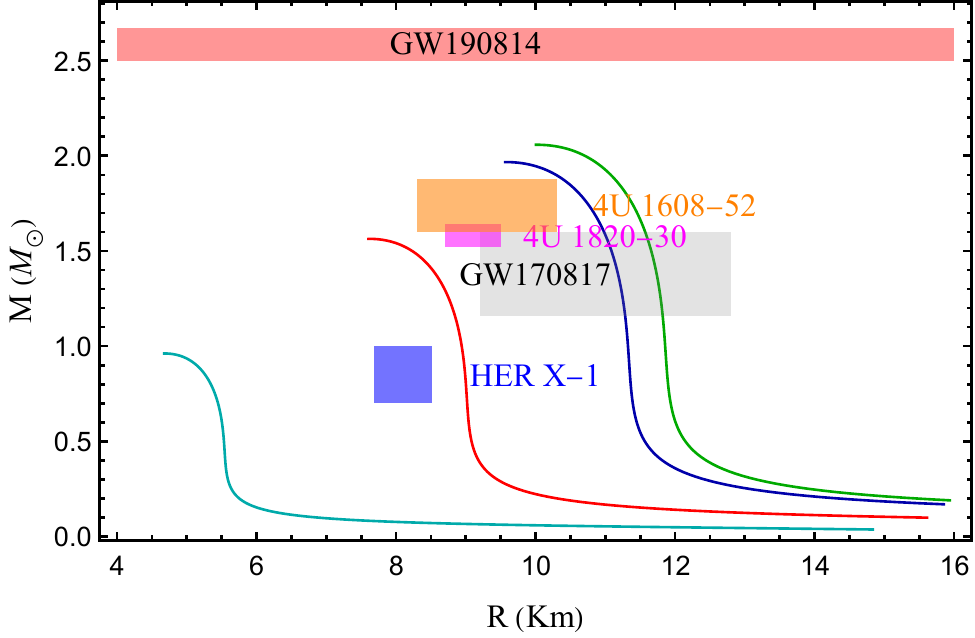}
		\subcaption{Maximum mass-radius plots using SLy4 EoS.}
		\label{fig3} 
	\end{minipage}
	\hfil
	\begin{minipage}{0.45\textwidth}
		\hspace{-0.5cm}
		\includegraphics[width=8cm]{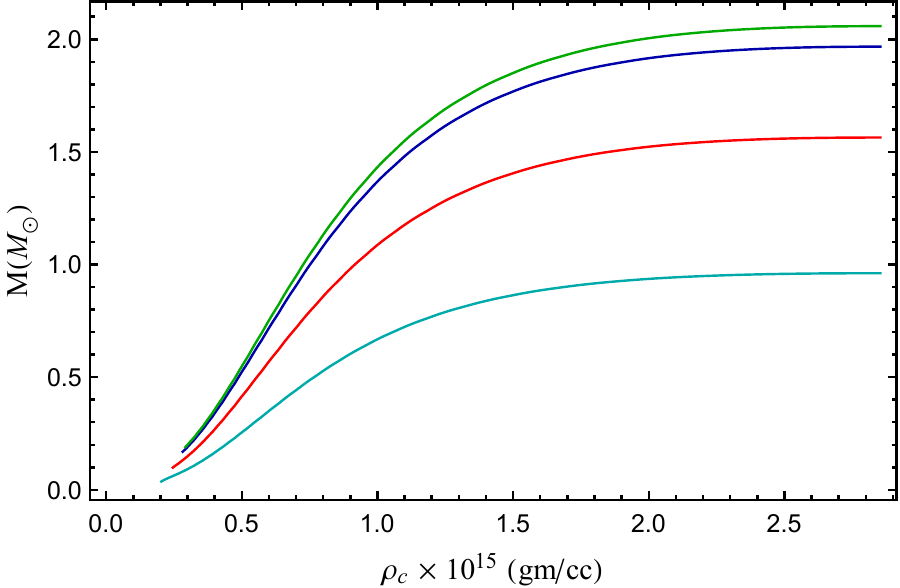}
		\subcaption{Mass vs Central density $(\rho_{c})$ using SLy4 EoS.}
		\label{fig4} 
	\end{minipage}
	\caption{The maximum mass-radius and mass vs central density diagrams for the Born-Infeld model using realistic neutron star EoSs, DD2 and SLy4, respectively. Here, (i) in the DD2 case, we have considered $Q_{0}=10^{-5}$ (Green), $Q_{0}=10^{-4}$ (Cyan), $Q_{0}=10^{-3}$ (Red) and $Q_{0}=0.1$ (Blue), and (ii) in the SLy4 case, we have considered $Q_{0}=10^{-5}$ (Cyan), $Q_{0}=10^{-4}$ (Red), $Q_{0}=10^{-3}$ (Blue) and $Q_{0}=0.1$ (Green).}
	\label{fig1a}
\end{figure}

In Fig.~\ref{fig1}, we demonstrate the maximum mass-radius plots for different values of $Q_{0}$. We have noted that with increasing $Q_{0}$, the maximum mass increases; this may be attributed to the fact that since DD2 and SLy4 remain the stiffest EoSs, with increasing $Q_{0}$, the term inside the square root approaches unity. Hence, the $f(Q)$ contributions cease to exist. Hence, we approach the GR limit of DD2 and SLy4 EoSs. Notably, beyond $Q_{0}=0.1$, the maximum mass saturates at $2.426~\mathrm{M_{\odot}}$ and $2.06~\mathrm{M_{\odot}}$ for DD2 and SLy4 cases respectively. Due to the novelty of the model, no physical bounds of $Q_{0}$ are present in the literature. Hence, we have considered $Q_{0}>0$ in this case. Furthermore, the Harrison-Zel'dovich-Novikov (HZN) condition \cite{Zeldovich}, demonstrating the variation of mass with central density $(\rho_{c})$, is illustrated in Fig.~\ref{fig2}. The maximum mass-radius values are tabulated in Table~\ref{tab1}. 

\begin{table}[htbp]
	\begin{minipage}{0.45\textwidth}
		\centering
		\begin{tabular}{ccc}
			\hline
			$Q_{0}$ & $M~\mathrm{(M_{\odot})}$ & $R$ (Km)\\ 
			\hline
			10$^{-5}$ & 1.13 & 5.57 \\
			10$^{-4}$ & 1.84 & 9.03 \\
			0.001 & 2.32 & 11.39 \\
			0.1 & 2.42 & 11.91 \\
			\hline
		\end{tabular} 
		\subcaption{}
		\label{tab1}
	\end{minipage}
	\hfil
	\begin{minipage}{0.45\textwidth}
		\centering
		\begin{tabular}{ccc}
			\hline
			$Q_{0}$ & $M~\mathrm{(M_{\odot})}$ & $R$ (Km)\\ 
			\hline
			$10^{-5}$ & 0.96 & 4.67 \\
			10$^{-4}$ & 1.56 & 7.60 \\
			10$^{-3}$ & 1.97 & 9.56 \\
			0.1 & 2.06 & 10.00 \\
			\hline
		\end{tabular}  
		\subcaption{}
		\label{tab2}
	\end{minipage}
	\caption{Tabulation of the maximum mass and corresponding radius obtained from Fig.~\ref{fig1a} for the parameter space considered in this study. Panel (a) presents the results obtained from Fig.~\ref{fig1} corresponding to the DD2 EoS, while panel (b) presents the results obtained from Fig.~\ref{fig3} corresponding to the SLy4 EoS.}
	\label{tab1a}
\end{table}

\subsection{Hu-Sawicki analogue: $f(Q)=Q-\mu\,Q_{0}\frac{(Q/Q_{0})^{m}}{1+(Q/Q_{0})^{m}}$}
The Hu-Sawicki model \cite{Hu} was proposed as a metric-variation model to describe an accelerated expansion of the universe without requiring the cosmological constant, while satisfying the solar system and cosmological tests in the small-field parametric limit. This model was entirely formulated to address the following theoretical and observational properties relevant to cosmology: (i) at high redshift, the formalism must replicate $\Lambda$CDM model, where it is strongly constrained by observations of CMB, (ii) in the low-redshift regime, it should accelerate the cosmic expansion without the introduction of a cosmological constant, (iii) To identify and validate small deviations from GR using the cosmological and solar system tests, it must retain the $\Lambda$CDM model in the limiting case, and so on. Following the cosmological framework, we introduce a Hu-Sawicki analogue that incorporates non-metricity and is expressed in the form:
\begin{equation}
	f(Q)=Q-\mu\,Q_{0}\frac{(Q/Q_{0})^{m}}{1+(Q/Q_{0})^{m}}, \label{eq3.2}
\end{equation}
where, $Q_{0}$ is the dimensionless the control parameter. Following the parameter space description of Hu-Sawicki model \cite{Hu}, we have considered $m>0$, and have also noted that the range of $\mu$ can be determined by $f_{Q}>0$ and $f_{QQ}>0$. However, typically $\mu$ ranges between 0.1 and 10. 

\begin{figure}[t!]
	\begin{minipage}{0.45\textwidth}
		\hspace{-1cm}
		\includegraphics[width=8cm]{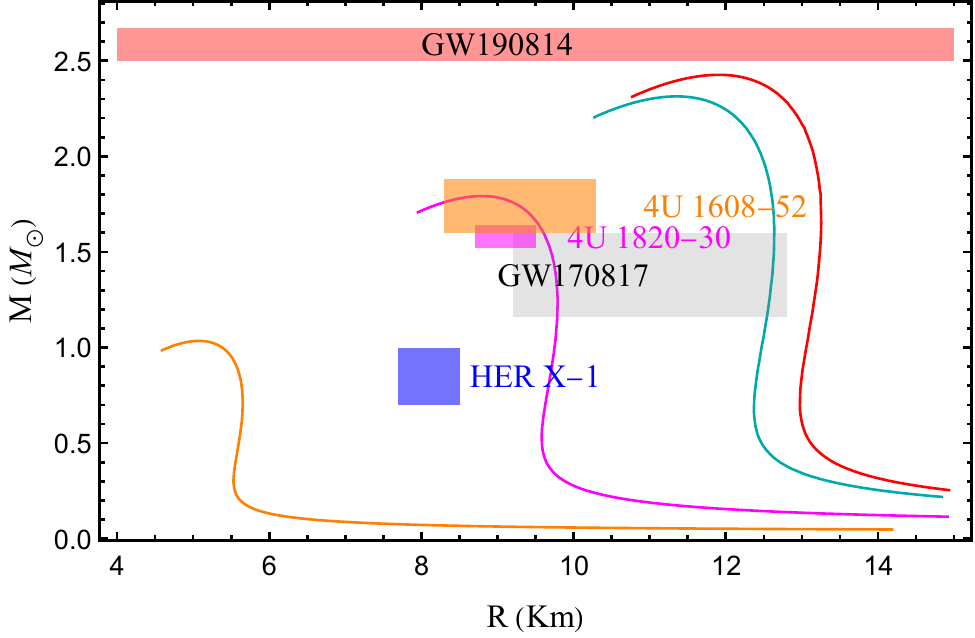}
		\subcaption{Maximum mass-radius plots using DD2 EoS.}
		\label{fig5} 
	\end{minipage}
	\hfil
	\begin{minipage}{0.45\textwidth}
		\hspace{-0.5cm}
		\includegraphics[width=8cm]{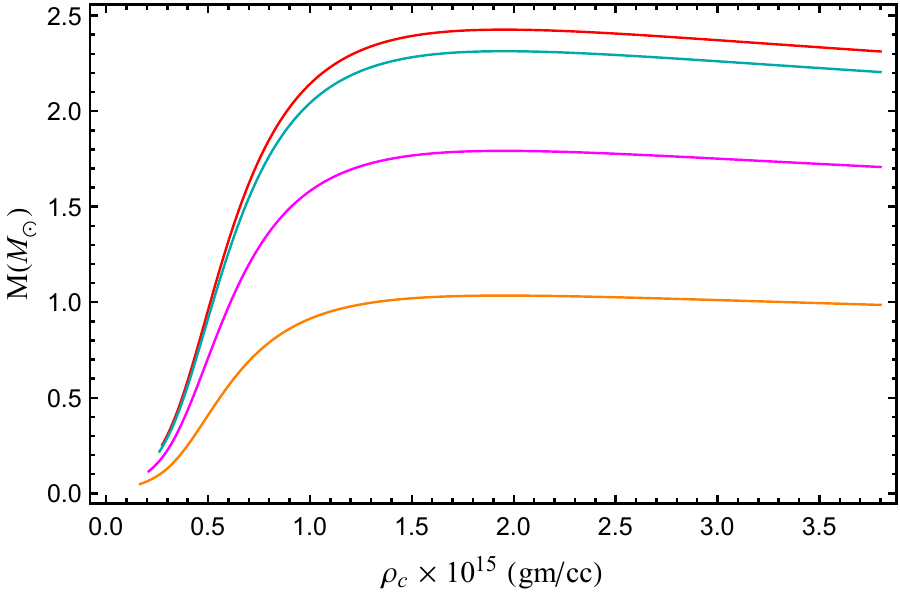}
		\subcaption{Mass vs Central density $(\rho_{c})$ plots using DD2 EoS.}
		\label{fig6} 
	\end{minipage}
	\begin{minipage}{0.45\textwidth}
		\hspace{-1cm}
		\includegraphics[width=8cm]{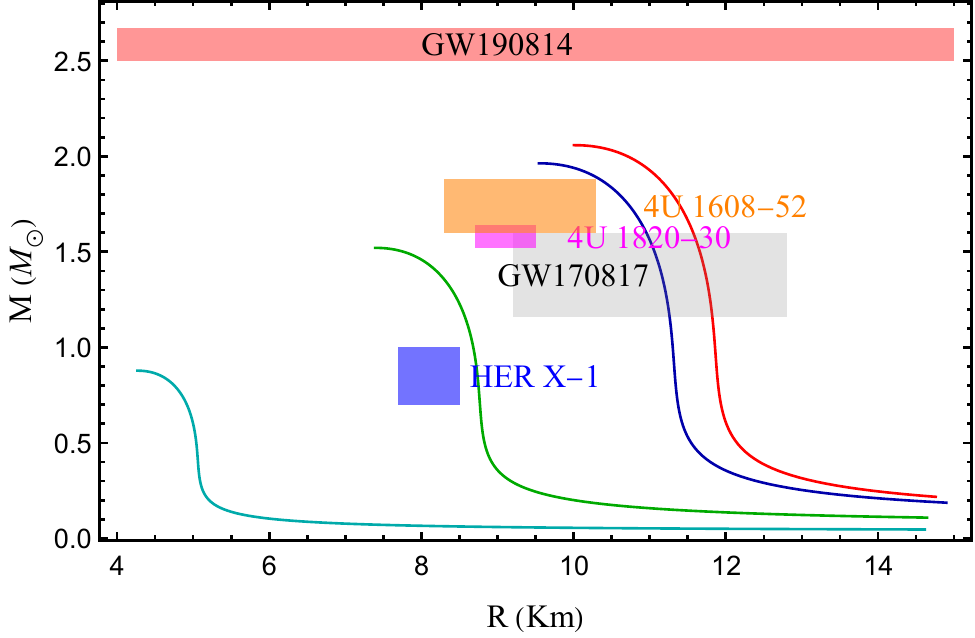}
		\subcaption{Maximum mass-radius plots using SLy4 EoS.}
		\label{fig7} 
	\end{minipage}
	\hfil
	\begin{minipage}{0.45\textwidth}
		\hspace{-0.5cm}
		\includegraphics[width=8cm]{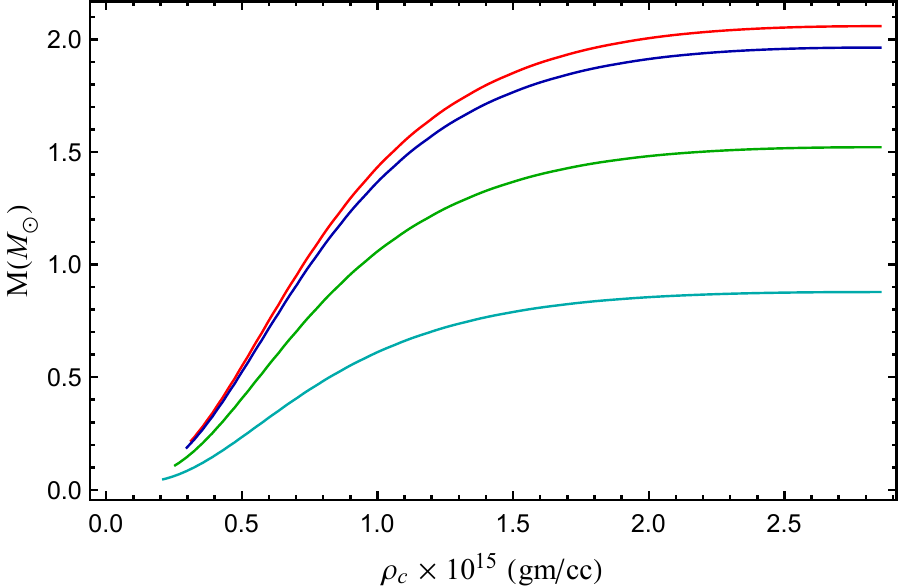}
		\subcaption{Mass vs Central density $(\rho_{c})$ plots using SLy4 EoS.}
		\label{fig8} 
	\end{minipage}
	\caption{Mass-radius and mass vs central density diagrams for Hu-Sawicki model using realistic neutron star EoS, DD2 and SLy4 for $m=1$. Here, (i) in the DD2 case, we have considered $\mu=10^{-7}$ (Red), $10^{-3}$ (Green), $0.1$ (Cyan), $0.5$ (Magenta) and  $0.9$ (Orange), and (ii) in the SLy4 case, we have considered $\mu=10^{-3}$ (Red), $0.1$ (Blue), $0.5$ (Green) and $0.9$ (Cyan).}
	\label{fig2a}
\end{figure}

We have studied the maximum mass-radius relations for realistic neutron stars using the Hu-Sawicki analogue, as expressed in Eq.~\ref{eq3.2}, and the corresponding results are illustrated in Fig.~\ref{fig2a} and are also tabulated in Table~\ref{tab2a}.
%%%%%%%%%%%%%%%%%%%%%%%%%%%%%%%%%%%%%%%%%%%
%%%%%%%%%%%%%%%%%%%%%%%%%%%%%%%%%%%%%%%%%%%
\begin{figure}[t!]
	\begin{minipage}{0.45\textwidth}
		\hspace{-1cm}
		\includegraphics[width=8cm]{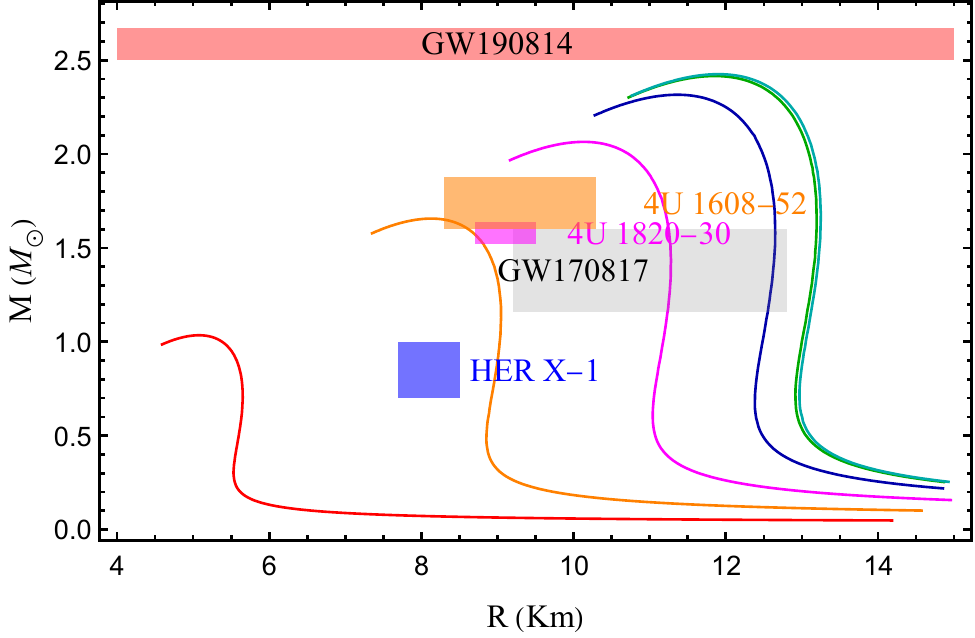}
		\subcaption{Maximum mass-radius plots using DD2 EoS.}
		\label{fig9} 
	\end{minipage}
	\hfil
	\begin{minipage}{0.45\textwidth}
		\hspace{-0.5cm}
		\includegraphics[width=8cm]{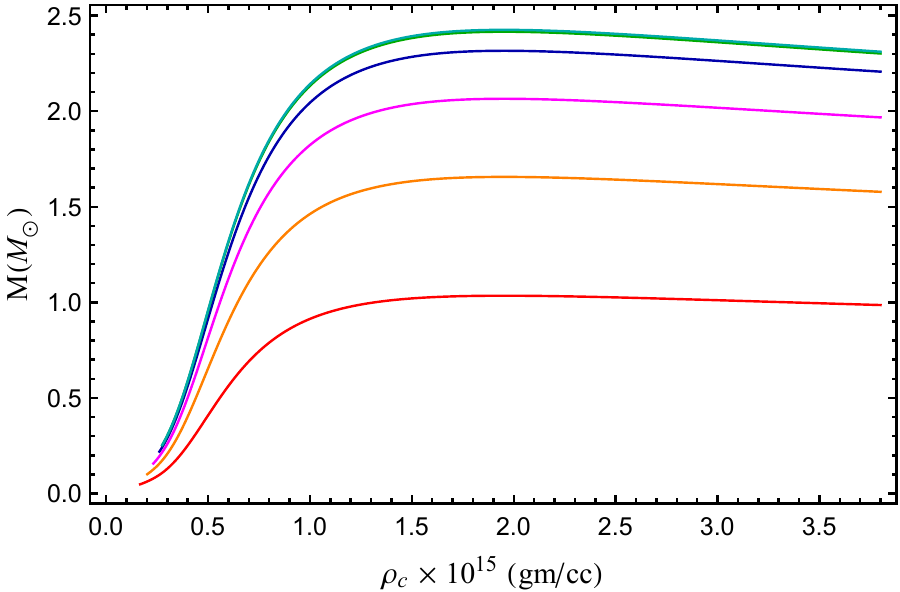}
		\subcaption{Mass vs Central density $(\rho_{c})$ plots using DD2 EoS.}
		\label{fig10} 
	\end{minipage}
	\begin{minipage}{0.45\textwidth}
		\hspace{-1cm}
		\includegraphics[width=8cm]{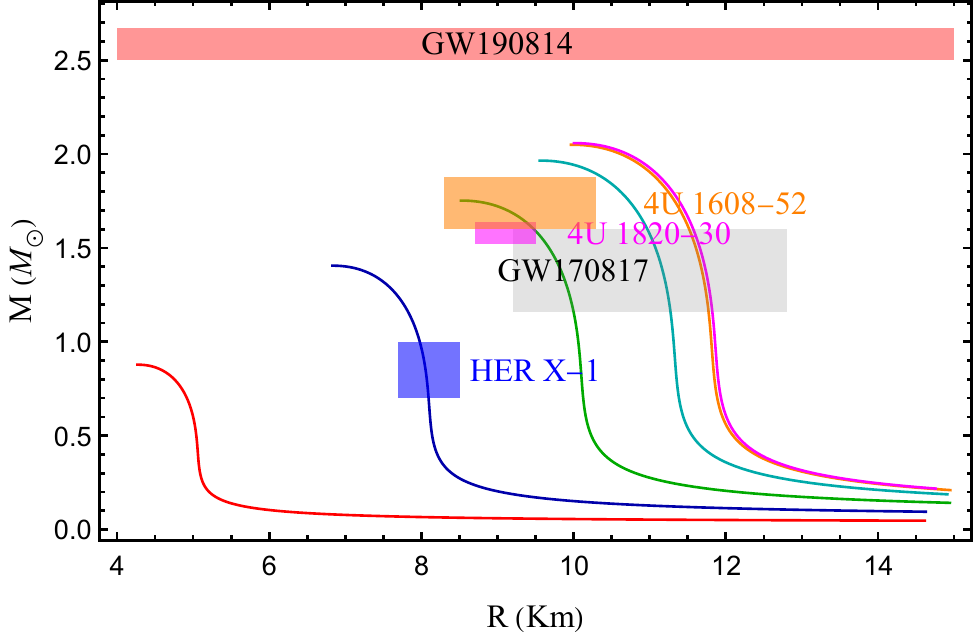}
		\subcaption{Maximum mass-radius plots using SLy4 EoS.}
		\label{fig11} 
	\end{minipage}
	\hfil
	\begin{minipage}{0.45\textwidth}
		\hspace{-0.5cm}
		\includegraphics[width=8cm]{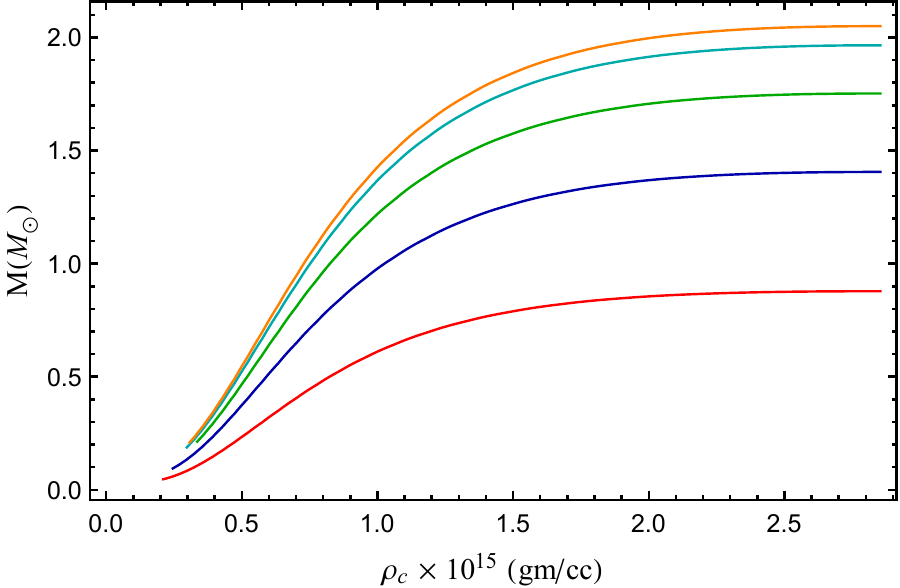}
		\subcaption{Mass vs Central density $(\rho_{c})$ plots using SLy4 EoS.}
		\label{fig12} 
	\end{minipage}
	\caption{Mass-radius and mass vs central density diagrams for Hu-Sawicki model using realistic neutron star EoS, DD2 and SLy4 for $\mu=0.9$. Here, (i) in the DD2 case, we have considered $m=1$ (Red), $1.2$ (Orange), $1.5$ (Magenta), $2$ (Blue), $3$ (Green) and $4$ (Cyan), and (ii) in the SLy4 case, we have considered $m=1$ (Red), $1.2$ (Blue), $1.5$ (Green), $2$ (Cyan), $3$ (Orange) and $4$ (Magenta).}
	\label{fig3a}
\end{figure}

\begin{table}[h!]
	\begin{minipage}{0.45\textwidth}
		\centering
		\begin{tabular}{ccc}
			\hline
			$\mu$ & $M~\mathrm{(M_{\odot})}$ & $R$ (Km)\\ 
			\hline
			$10^{-3}$ & 2.42 & 11.91 \\
			0.1 & 2.31 & 11.36 \\
			0.5 & 1.79 & 8.80 \\
			0.9 & 1.03 & 5.08 \\
			\hline
		\end{tabular}
		\subcaption{}
		\label{tab3}
	\end{minipage}
	\hfil
	\begin{minipage}{0.45\textwidth}
		\centering
		\begin{tabular}{ccc}
			\hline
			$\mu$ & $M~\mathrm{(M_{\odot})}$ & $R$ (Km)\\ 
			\hline
			$10^{-3}$ & 2.058 & 10.00 \\
			0.1 & 1.96 & 9.54 \\
			0.5 & 1.52 & 7.39 \\
			0.9 & 0.88 & 4.27 \\
			\hline
		\end{tabular}
		\subcaption{}
		\label{tab4}
	\end{minipage}
	\caption{Tabulation of the maximum mass and corresponding radius obtained from Fig.~\ref{fig2a} for the parameter space considered in this study. Panel (a) presents the results obtained from Fig.~\ref{fig5} corresponding to the DD2 EoS, while panel (b) presents the results obtained from Fig.~\ref{fig7} corresponding to the SLy4 EoS.}
	\label{tab2a}
\end{table}
%%%%%%%%%%%%%%%%%%%%%%%%%%%%%%%%%%%%%%%%%%%%
%%%%%%%%%%%%%%%%%%%%%%%%%%%%%%%%%%%%%%%%%%%%
\begin{table}[h!]
	\begin{minipage}{0.45\textwidth}
		\centering
		\begin{tabular}{ccc}
			\hline
			$m$ & $M~\mathrm{(M_{\odot})}$ & $R$ (Km)\\ 
			\hline
			1 & 1.03 & 5.08 \\
			1.2 & 1.65 & 8.13 \\
			1.5 & 2.06 & 10.14 \\
			2 & 2.31 & 11.37 \\
			3 & 2.41 & 11.86 \\
			4 & 2.42 & 11.91 \\
			\hline
		\end{tabular}
		\subcaption{}
		\label{tab5}
	\end{minipage}
	\hfil
	\begin{minipage}{0.45\textwidth}
		\centering
		\begin{tabular}{ccc}
			\hline
			$m$ & $M~\mathrm{(M_{\odot})}$ & $R$ (Km)\\ 
			\hline
			1 & 0.88 & 4.27 \\
			1.2 & 1.40 & 6.83 \\
			1.5 & 1.75 & 8.52 \\
			2 & 1.96 & 9.55 \\
			3 & 2.05 & 9.96 \\
			4 & 2.06 & 10.00 \\
			\hline
		\end{tabular}
		\subcaption{}
		\label{tab6}
	\end{minipage}
	\caption{Tabulation of the maximum mass and corresponding radius obtained from Fig.~\ref{fig3a} for the parameter space considered in this study. Panel (a) presents the results obtained from Fig.~\ref{fig9} corresponding to the DD2 EoS, while panel (b) presents the results obtained from Fig.~\ref{fig11} corresponding to the SLy4 EoS.}
	\label{tab3a}
\end{table}

In the present scenario, we have obtained the maximum mass-radius relations, using DD2 and SLy4 EoSs, for two distinct parameter spaces: (i) Fig.~\ref{fig2a}, corresponding to $m=1$ with varying $\mu$, and (ii) Fig.~\ref{fig3a}, corresponding to $\mu=0.9$ with varying $m$. From Table~\ref{tab2a} associated with Fig.~\ref{fig2a}, we note that for a particular choice of $Q/Q_{0}$ and $m$, the maximum mass decreases with increasing $\mu$. This may be attributed to the fact that, with increasing $\mu$, the correction to the $f(Q)$ model becomes stronger, which in turn reduces the magnitude of the entire $f(Q)$ model. Consequently, the deviation from the GR case $(f(Q)=Q)$ becomes prominent. These effects together may alter the gravitational dynamics, thereby reducing the maximum mass-radius. On the other hand, from Table~\ref{tab3a} associated with Fig.~\ref{fig3a}, we observe that, for fixed values of $Q/Q_{0}$ and $\mu$, the maximum mass increases with increasing $m$. This may be explained in the following way: the parameter $m$ controls how quickly the modifications to the gravitational sector switch on. Within the present parameter space, increasing $m$ suppresses the modifications of the $f(Q)$ model, and it approaches the GR limit, $f(Q)=Q$. This weakening of the modifications allows equilibrium configurations to support more mass, as evident from Fig.~\ref{fig3a}. Additionally, since $m>0$ in this model, fractional powers are also allowed, as tabulated in Table~\ref{tab3a}. Regarding the parameter bounds, we note that for a fixed value of $m$, the obtained maximum masses approach their respective upper limits, {\it viz.} 2.426 $\mathrm{M_{\odot}}$ for DD2 and 2.06 $\mathrm{M_{\odot}}$ for SLy4, when $\mu<10^{-3}$ for both the EoSs. Furthermore, the HZN conditions for both cases are illustrated in Figs.~\ref{fig2a} and \ref{fig3a}, describing physically realistic variations.

\subsection{Tsujikawa-inspired model: $f(Q)=Q+\alpha\,Q_{0}\tanh{\left(\frac{Q}{Q_{0}}\right)}$}
\begin{figure}[t!]
	\begin{minipage}{0.45\textwidth}
		\hspace{-1cm}
		\includegraphics[width=8cm]{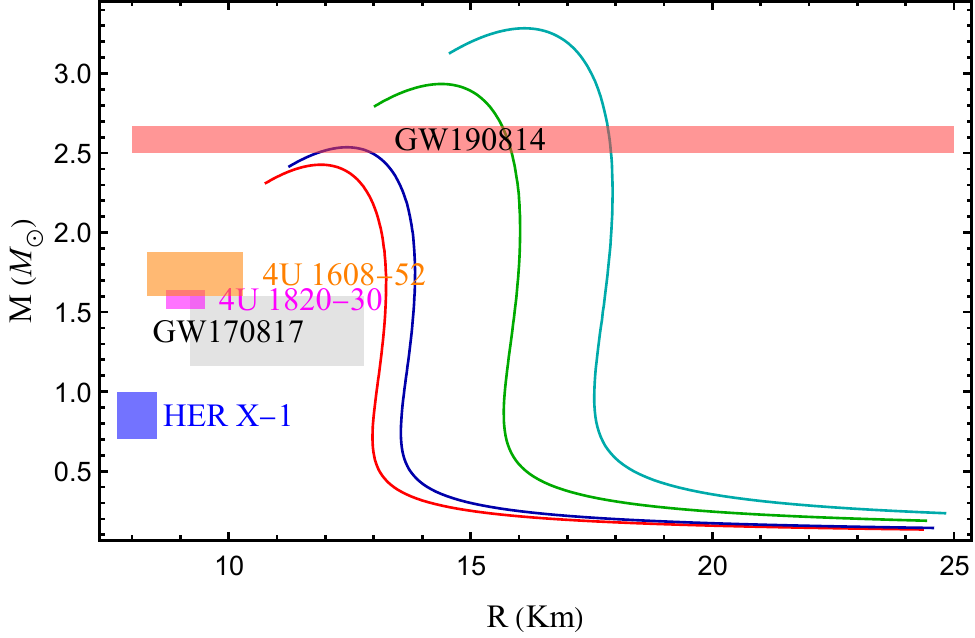}
		\subcaption{Maximum mass-radius plots using DD2 EoS.}
		\label{fig13} 
	\end{minipage}
	\hfil
	\begin{minipage}{0.45\textwidth}
		\hspace{-0.5cm}
		\includegraphics[width=8cm]{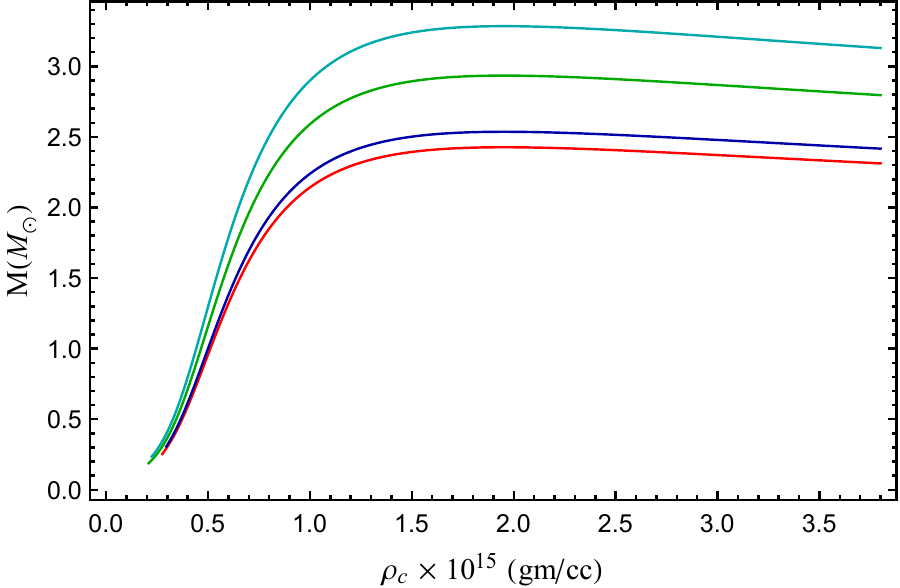}
		\subcaption{Mass vs Central density $(\rho_{c})$ plots using DD2 EoS.}
		\label{fig14} 
	\end{minipage}
	\begin{minipage}{0.45\textwidth}
		\hspace{-1cm}
		\includegraphics[width=8cm]{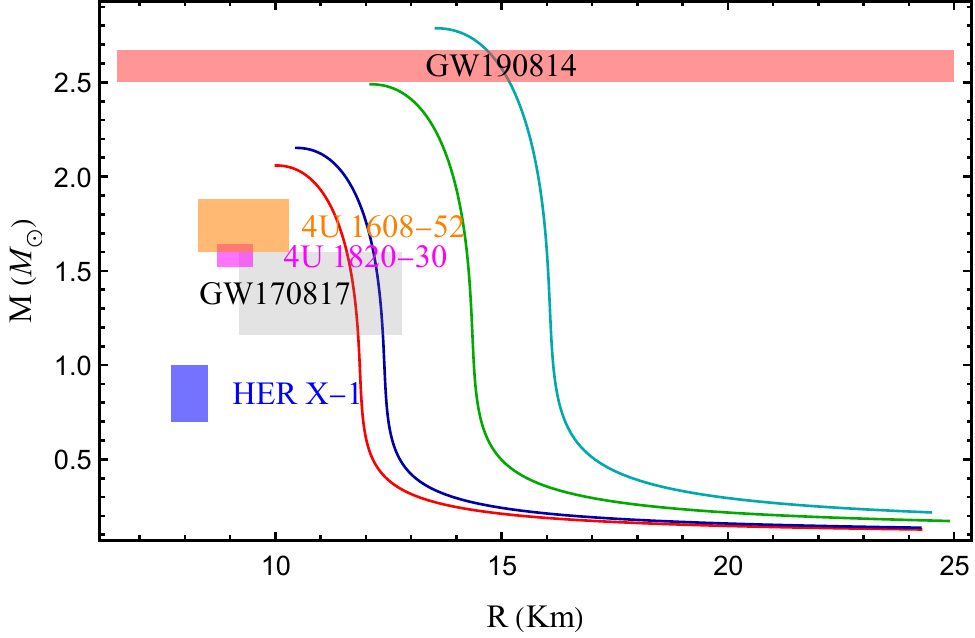}
		\subcaption{Maximum mass-radius plots using SLy4 EoS.}
		\label{fig15} 
	\end{minipage}
	\hfil
	\begin{minipage}{0.45\textwidth}
		\hspace{-0.5cm}
		\includegraphics[width=8cm]{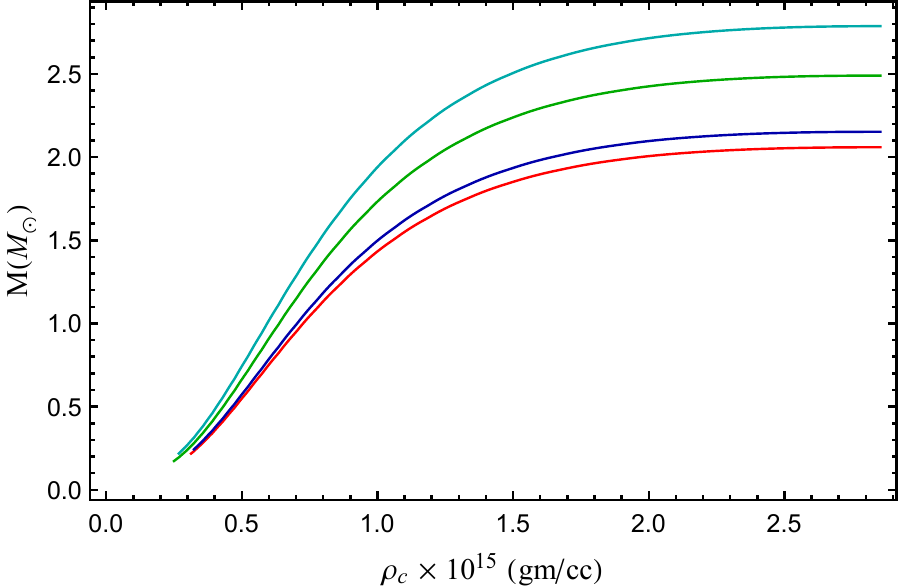}
		\subcaption{Mass vs Central density $(\rho_{c})$ plots using SLy4 EoS.}
		\label{fig16} 
	\end{minipage}
	\caption{Mass-radius and mass vs central density diagrams for Tsujikawa model using realistic neutron star EoS, DD2 and SLy4 for different values of $\alpha$. In both the cases, we have considered $\alpha=0$ (Red), $0.1$ (Blue), $0.5$ (Green) and $0.9$ (Cyan).}
	\label{fig4a}
\end{figure}

The Tsujikawa model \cite{Tsujikawa} was proposed to discuss a cosmological dark energy scenario that satisfies cosmological viability and local gravity tests. Notably, this model provides a more general framework of the Hu-Sawicki and Starobinsky models \cite{Starobinsky}. Following the detailed mathematical formalism described in Ref.~\cite{Tsujikawa}, we have expressed the $f(Q)$ model in the form:
\begin{equation}
	f(Q)=Q+\alpha\,Q_{0}\tanh{\left(\frac{Q}{Q_{0}}\right)}, \label{eq3.3}
\end{equation}
where $\alpha$ is a characteristic parameter that measures the magnitude of deviation from GR. In this model, we have considered $0\leq\alpha\leq1$ \cite{Tsujikawa}. Within this parameter space, the obtained mass-radius results are demonstrated in Fig.~\ref{fig4a}.

The mass-radius values extracted from Fig.~\ref{fig4a} are tabulated in Table~\ref{tab4a} for both the EoSs. From Table~\ref{tab4a}, it is evident that the maximum mass increases with increasing $\alpha$. This may be explained in the following way: with increasing $\alpha$, the corrections to the $f(Q)$ model becomes more prominent, thereby altering the gravitational dynamics of the stellar configuration. Consequently, these corrections may provide additional pressure support against gravitational collapse, and hence the configuration can support a more massive structure. Moreover, to obtain the optimum mass-radius values, we have considered $\alpha=0.99$ and obtained $M=3.36~\mathrm{M_{\odot}}$ for the DD2 case and $M=2.85~\mathrm{M_{\odot}}$ for the SLy4 case. We observe that the maximum mass-radius values obtained in the Tsujikawa model, using the DD2 and SLy4 EoSs, are significantly higher than their GR counterparts. Consequently, the recently observed high-mass compact stars, including the secondary component of GW190814 with a mass of $2.59^{+0.08}_{-0.09}~\mathrm{M_{\odot}}$ \cite{Abbott}, can be naturally accommodated within the Tsujikawa model.    
\begin{table}[h!]
	\begin{minipage}{0.45\textwidth}
		\centering
		\begin{tabular}{ccc}
			\hline
			$\alpha$ & $M~\mathrm{(M_{\odot})}$ & $R$ (Km)\\ 
			\hline
			0 & 2.42 & 11.92 \\
			0.1 & 2.54 & 12.46 \\
			0.5 & 2.93 & 14.41 \\
			0.9 & 3.28 & 16.13 \\
			\hline
		\end{tabular} 
		\subcaption{}
		\label{tab7}
	\end{minipage}
	\hfil
	\begin{minipage}{0.45\textwidth}
		\centering
		\begin{tabular}{ccc}
			\hline
			$\alpha$ & $M~\mathrm{(M_{\odot})}$ & $R$ (Km)\\ 
			\hline
			0 & 2.06 & 10.00 \\
			0.1 & 2.15 & 10.46 \\
			0.5 & 2.49 & 12.10 \\
			0.9 & 2.78 & 13.55 \\
			\hline
		\end{tabular}  
		\subcaption{}
		\label{tab8}
	\end{minipage}
	\caption{Tabulation of the maximum mass and corresponding radius obtained from Fig.~\ref{fig4a} for the parameter space considered in this study. Panel (a) presents the results obtained from Fig.~\ref{fig13} corresponding to the DD2 EoS, while panel (b) presents the results obtained from Fig.~\ref{fig15} corresponding to the SLy4 EoS.}
	\label{tab4a}
\end{table}  

\section{Assessing physical viability via radius prediction}\label{sec4}
To scrutinise the consistency of the introduced $f(Q)$ models, we have predicted the radii of a number of recently observed compact stars and the low-mass companions of the gravitational wave events from the TOV solutions for the three cases. The results are summarised in Tables~\ref{tab9}, \ref{tab10} and \ref{tab11}.
\begin{table}[t!]
	\centering
	\resizebox{\textwidth}{!}{$
		\begin{tabular}{ccccccc}
			\hline
			\multirow{3}{*}{Compact stars} & \multirow{3}{*}{Measured mass} & \multirow{3}{*}{Measured radius} & \multicolumn{2}{c}{DD2} & \multicolumn{2}{c}{SLy4}\\ \cline{4-7}
			&  & & \multirow{2}{*}{$Q_{0}\times10^{-4}$} & Predicted radius & \multirow{2}{*}{$Q_{0}\times10^{-4}$} & Predicted radius \\ 
			& $(\mathrm{M_{\odot}})$ & (Km) && (Km) & & (Km) \\
			\hline
			HER X-1 \cite{Abubekerov} & $0.85\pm0.15$ & $8.1\pm0.41$ & $0.33$ & 8.1 & 0.57 & 8.1 \\
			\vspace{0.1cm}		
			EXO 1745-248 \cite{Ozel} & 1.4 & 11 & 1.8 & 11 & 7.5 & 11\\
			\vspace{0.1cm}
			4U 1820-30 \cite{Guver} & $1.58\pm0.06$ & $9.1\pm0.4$ & 0.66 & 9.1 & 1.7 & 9.07\\
			\vspace{0.1cm}
			4U 1608-52 \cite{Guver1} & $1.74\pm0.14$ & $9.3\pm1.0$ & 0.85 & 9.3 & 2.7 & 9.31 \\
			\vspace{0.1cm}
			GW170817 \cite{Abbott1} & 1.16-1.60 & $10.7^{+2.1}_{-1.5}$ & 1.5 & $10.67^{+0.02}_{-0.02}$ & 5.5 & $10.7^{+0.18}_{-0.17}$ \\
			\vspace{0.1cm}
			GW190814 \cite{Abbott} & $2.59^{+0.08}_{-0.09}$ & -- & -- & -- &&\\ \hline
		\end{tabular}$}
	\caption{Predicted radius using Born-Infeld inspired model for parametric choices of $Q_{0}$.}
	\label{tab9}
\end{table}

\begin{table}[t!]
	\centering
	\resizebox{\textwidth}{!}{$
		\begin{tabular}{ccccccccc}
			\hline
			\multirow{3}{*}{Compact stars} & \multirow{3}{*}{Measured mass} & \multirow{3}{*}{Measured radius} & \multicolumn{3}{c}{DD2} & \multicolumn{3}{c}{SLy4}\\ \cline{4-9}
			&  & & \multirow{2}{*}{$\mu$} & \multirow{2}{*}{m} & Predicted radius & \multirow{2}{*}{$\mu$} & \multirow{2}{*}{m} & Predicted radius \\
			& $(\mathrm{M_{\odot}})$ & (Km) & & & (Km) & & & (Km) \\
			\hline
			HER X-1 \cite{Abubekerov} & $0.85\pm0.15$ & $8.1\pm0.41$ & 0.5 & 0.85 & 8.12 & 0.9 & 1.21 & 8.15 \\
			\vspace{0.1cm}		
			EXO 1745-248 \cite{Ozel} & 1.4 & 11 & 0.9 & 1.45 & 11.02 &0.8 & 1.85 & 11 \\
			\vspace{0.1cm}
			4U 1820-30 \cite{Guver} & $1.58\pm0.06$ & $9.1\pm0.4$ & 1.1 & 1.32 & 9.09 & 069 & 1.31 & 9.07\\ 
			\vspace{0.1cm}
			4U 1608-52 \cite{Guver1} & $1.74\pm0.14$ & $9.3\pm1.0$ & 0.9 & 1.28 & 9.36 & 0.5 & 1.29 & 9.30 \\
			\vspace{0.1cm}
			GW170817 \cite{Abbott1} & 1.16-1.60 & $10.7^{+2.1}_{-1.5}$ & 0.88 & 1.4 & $10.74^{+0.02}_{-0.02}$ & 0.5 & 1.52 & $10.71^{+0.18}_{-0.17}$ \\
			\vspace{0.1cm}
			GW190814 \cite{Abbott} & $2.59^{+0.08}_{-0.09}$ & -- & -- & -- & -- & -- & -- & --\\
			\hline
		\end{tabular}$}
	\caption{Predicted radius using Hu-Sawicki analogue for parametric choices of $\mu$ and $m$.}
	\label{tab10}
\end{table}

\begin{table}[t!]
	\centering
	\resizebox{\textwidth}{!}{$
		\begin{tabular}{ccccccc}
			\hline
			\multirow{3}{*}{Compact stars} & \multirow{3}{*}{Measured mass} & \multirow{3}{*}{Measured radius} & \multicolumn{2}{c}{DD2} & \multicolumn{2}{c}{SLy4}\\ \cline{4-7}
			&  & & \multirow{2}{*}{$\alpha$} & Predicted radius & \multirow{2}{*}{$\alpha$} & Predicted radius \\ 
			& $(\mathrm{M_{\odot}})$ & (Km) && (Km) & & (Km) \\
			\hline
			HER X-1 \cite{Abubekerov} & $0.85\pm0.15$ & $8.1\pm0.41$ & 0.1 & 13.57 & 0.1 & 12.40\\
			\vspace{0.1cm}		
			EXO 1745-248 \cite{Ozel} & 1.4 & 11 & 0.1 & 13.79 & 0.1 & 12.29 \\
			\vspace{0.1cm}
			4U 1820-30 \cite{Guver} & $1.58\pm0.06$ & $9.1\pm0.4$ & 0.1 & 13.81 & 0.1 & 12.18\\
			\vspace{0.1cm}
			4U 1608-52 \cite{Guver1} & $1.74\pm0.14$ & $9.3\pm1.0$ & 0.1 & 13.85 & 0.1 & 12.02\\
			\vspace{0.1cm}
			GW170817 \cite{Abbott1} & 1.16-1.60 & $10.7^{+2.1}_{-1.5}$ & 0.1 & $13.77^{+0.07}_{-0.07}$ & 0.1 & $12.26^{+0.11}_{-0.10}$\\
			\vspace{0.1cm}
			GW190814 \cite{Abbott} & $2.59^{+0.08}_{-0.09}$ & -- & 0.15 & 12.72 & 0.63 & 12.59 \\ \hline
		\end{tabular}$}
	\caption{Predicted radius using Tsujikawa-inspired model for parametric choices of $\alpha$.}
	\label{tab11}
\end{table}

From Tables~\ref{tab9}, \ref{tab10} and \ref{tab11}, we note the following: 
\begin{itemize}
	\item For low-mass $(<2\,\mathrm{M_{\odot}})$ neutron stars/compact stars, the predicted radii are in good agreement with estimates from recent observations for the proposed Born-Infeld and Hu-Sawicki models admitting both the DD2 and SLy4 EoSs. 
	\item Since the maximum masses obtained with DD2 and SLy4 EoSs in both Born-Infeld and Hu-Sawicki models are capped at $<2.59\,\mathrm{M_{\odot}}$, explanation of the secondary object in GW190814 with an estimated mass of $2.59^{+0.08}_{-0.09}\,\mathrm{M_{\odot}}$ is impossible for these cases, within the parameter space used here. 
	\item For the Tsujikawa-inspired model, the obtained maximum masses for both DD2 and SLy4 EoSs are sufficiently large to accommodate the GW190814 secondary. Hence, on the one hand, the Tsujikawa-inspired model can be implemented to explain the GW190814 secondary, while on the other hand, for the low-mass $(<2\,\mathrm{M_{\odot}})$ neutron stars/compact stars, the predicted radii based on this model are not compatible with the estimates inferred from recent observations. 
	\item In the case of Tsujikawa-inspired model, we have noted that increasing $\alpha$ increases the predicted radii to a large extent. Hence, to show the impact of the proposed model within this constant-Q branch of $f(Q)$ gravity, we have considered $\alpha=0.1$, representatively. However, for GW190814, we predict the radius by fine-tuning $\alpha$.    
	\item On the basis of the above analysis, it may be claimed that, for the realistic EoSs used here, the Born-Infeld and Hu-Sawicki models are much more plausible for low-mass neutron stars/compact stars, whereas the Tsujikawa-inspired model is physically feasible for $>2\,\mathrm{M_{\odot}}$.  
\end{itemize}

\section{Discussion}\label{sec5}
In recent years, a genuine interest has grown in obtaining a beyond-GR theory of gravity where affine connections uphold their dynamical nature. In this context, $f(Q)$ gravity has garnered significant attention, particularly in the context of black holes and cosmology. While the non-metricity based theory of $f(Q)$ gravity has been exemplary in the above cases, investigating the compact stars within this framework often introduces simplistic assumptions that suppress the required degrees of freedom pertaining to $f(Q)$ gravity. These simplified constructions provide useful reference solutions and have contributed to mapping the phenomenology of compact objects within the theory. They may also implicitly retain gauge choices and structures characteristic of symmetric teleparallel equivalent of GR. Consequently, the resulting stellar configurations can exhibit only modest departures from their general relativistic counterparts, potentially obscuring genuinely theory-specific signatures of the underlying $f(Q)$ dynamics. 

In light of the above, we have considered the detailed analysis put forward by Heisenberg and Pastor-Marcos \cite{Heisenberg}. Starting from the metric-affine theory, we have obtained the relevant connection relations and the modified field equation in Eq.~(\ref{eq10}). Considering a static, spherically symmetric spacetime, and making use of the parametric relations presented in Eqs.~(\ref{eq15})-(\ref{eq24}), we derive the corresponding TOV equations, as given in Eqs.~(\ref{eq25})-(\ref{eq28}). By imposing regularity at the centre and the boundary constraints, we have noted that for different choices of $f(Q)$, the regularity naturally selects 
the branch which is characterised by a constant non-metricity. Accordingly, the TOV equations are modified to a simpler form. Notably, this branch consequently provides a controlled setting in which the effect of the functional form of $f(Q)$ can be examined while maintaining the regularity requirements of the underlying covariant formulation. 

In this framework, we have introduced and examined three novel nonlinear models, namely a Born-Infeld-inspired model \cite{Born,Born1}, $f(Q)=Q_{0}\left[\sqrt{1+\frac{2Q}{Q_{0}}}-1\right]$,  a Hu-Sawicki analogue \cite{Hu}, $f(Q)=Q-\mu Q_{0}\frac{(Q/Q_{0})^{m}}{1+(Q/Q_{0})^{m}}$, 
and a Tsujikawa-inspired model \cite{Tsujikawa}, $f(Q)=Q+\alpha Q_{0}\tanh\left(\frac{Q}{Q_{0}}\right).$ The modified TOV equations have been solved using two realistic unified neutron star EoS, namely GPPVA(DD2) \cite{Grill,Typel,Pearson} and SLy4 \cite{Gulminelli,Chabanat,Danielewicz}. We have examined the resulting mass-radius sequences, the mass-central density behaviour through the Harrison-Zel'dovich-Novikov criterion, and the radii associated with several observed compact stars.

For the Born-Infeld inspired model, increasing the control parameter $Q_{0}$ progressively weakens the departure from the GR configuration as evident in Fig~\ref{fig1a}. The maximum masses consequently approach the corresponding GR values of approximately $2.42,M_{\odot}$ for DD2 and $2.06,M_{\odot}$ for SLy4. The mass ranges are summarised in Table~\ref{tab1a}. A similar trend is obtained for the Hu-Sawicki analogue. For fixed $m$, increasing $\mu$ strengthens the modification of the gravitational sector and leads to a systematic reduction of the maximum mass. Conversely, at fixed $\mu$, increasing $m$ weakens the effective modification over the relevant range of $Q/Q_{0}$ and drives the solutions towards the GR limit. These natures are illustrated in Figs.~\ref{fig2a} and \ref{fig3a}, and the mass-radius values are tabulated in Tables~\ref{tab2a} and \ref{tab3a}. The Tsujikawa-inspired model exhibits a qualitatively different behaviour. Increasing $\alpha$ enhances the modification of the gravitational sector and increases the maximum mass supported by the stellar configurations. For $\alpha=0.9$, we obtain maximum masses of approximately $3.28,M_{\odot}$ and $2.78,M_{\odot}$ for DD2 and SLy4, respectively, while extending the parameter to $\alpha=0.99$ yields maximum masses of approximately $3.36,M_{\odot}$ and $2.85,M_{\odot}$. The results are summarised in Table~\ref{tab4a} associated with Fig.~\ref{fig4a}.  

It is also worth noting that all three models recover the linear form \(f(Q)=Q\) in their respective limiting cases: \(Q_0\rightarrow\infty\) for the Born-Infeld-inspired model, \(\mu\rightarrow0\) for the Hu-Sawicki analogue, and \(\alpha\rightarrow0\) for the Tsujikawa-inspired model. These limits provide the corresponding general relativistic reference configurations against which the modified stellar solutions can be compared.

To assess the physical viability of the present framework, we have predicted the radii of different compact stars for these models and the results are tabulated in Tables~\ref{tab9}, \ref{tab10} and \ref{tab11}. It is interesting to note that in the cases of Born-Infeld and Hu-Sawicki models, the resulting configurations reproduce the observed radii of several low-mass compact stars with good accuracy, for both DD2 and SLy4. Nevertheless, for these models, the maximum masses remain below the mass inferred for the GW190814 secondary $(2.59^{+0.08}_{-0.09})$ throughout the parameter space considered. Hence, these models, in their present constant-$Q$ realisations, are more naturally suited to describing ordinary low and intermediate-mass neutron stars than the high-mass secondary of GW190814. In case of Tsujikawa model, the maximum mass values are sufficiently large to accommodate a compact object with a mass comparable to the secondary component of GW190814. However, the same modification also produces substantially larger radii for low-mass stars than those inferred from the multi-messenger observational estimates. This highlights an important tension: within the constant-$Q$ branch, the parameter region that increases the maximum mass sufficiently to accommodate GW190814 does not simultaneously provide an equally satisfactory description of the radii of low-mass neutron stars.

The combined mass-radius analysis therefore indicates that the three models do not provide a uniform improvement over GR across the entire observed compact-star population. Rather, their phenomenological effects depend strongly on the form and parameters of $f(Q)$. The Born-Infeld-inspired and Hu-Sawicki analogues remain comparatively close to the GR predictions over part of their viable parameter spaces and provide reasonable descriptions of low-mass compact stars, but they do not reach the mass scale of GW190814 within the parameter ranges investigated. The Tsujikawa-inspired model can generate substantially more massive configurations, but this improvement comes at the cost of predicting comparatively large radii for low-mass objects. This complementarity between the maximum mass and radius constraints provides a useful way of discriminating among different functional forms of $f(Q)$.

It is important to emphasise that the present results concern specifically the regular constant-$Q$ branch of the covariant $f(Q)$ framework. Since $\partial_{r}Q=0$ on this branch, the explicit derivative contributions involving $f''(Q)$ and $f^{(3)}(Q)$ vanish from the stellar equations. Consequently, the results should not be interpreted as a complete exploration of the dynamical degrees of freedom of generic $f(Q)$ gravity. Instead, they establish how different nonlinear functional forms modify neutron star structure within a regular branch selected by the central boundary conditions. In this sense, the present analysis also demonstrates that even this restricted branch can produce appreciable changes in the maximum mass and radius, depending on the chosen functional form.

Overall, our results demonstrate that realistic neutron star EoS, when combined with the covariant constant-$Q$ formulation of $f(Q)$ gravity, provide a useful framework for confronting modified gravity with multimessenger compact star observations. In particular, the simultaneous use of maximum mass and radius constraints is essential, since a model capable of supporting very massive compact objects need not automatically reproduce the radii of ordinary neutron stars. A more complete assessment of the theory would require extending the present analysis to the genuinely non-constant-$Q$ branch, where the connection dynamics and the derivative terms involving $f_{Q}$, $f_{QQ}$ and higher derivatives contribute explicitly. Such an extension, together with tidal deformability, moment of inertia and gravitational wave constraints, would provide a more stringent test of the viability of nonlinear $f(Q)$ models in the strong-field regime.
\acknowledgments

DB is thankful to the Department of Science and Technology (DST), Govt. of India, for providing the fellowship vide no:  DST/INSPIRE Fellowship/2021/IF210761. PKC gratefully acknowledges support from IUCAA, Pune, India under Visiting Associateship programme.

%\paragraph{Note added.} This is also a good position for notes added
%after the paper has been written.

% Bibliography

%% [A] Recommended: using JHEP.bst file
%% \bibliographystyle{JHEP}
%% \bibliography{biblio.bib}

%% or
%% [B] Manual formatting (see below)
%% (i) We suggest to always provide author, title and journal data or doi:
%% in short all the informations that clearly identify a document.
%% (ii) please avoid comments such as "For a review'', "For some examples",
%% "and references therein" or move them in the text. In general, please leave only references in the bibliography and move all
%% accessory text in footnotes.
%% (iii) Also, please have only one work for each \bibitem.

\end{document}